\documentclass[aps,twocolumn]{revtex4-1}

\usepackage{graphicx}
\usepackage{bm}
\usepackage{braket}
\usepackage{hyperref}
\usepackage[usenames]{xcolor}
\usepackage{amsmath}
\usepackage{graphicx}
\usepackage{latexsym}
\usepackage{amsmath,amssymb}
\usepackage{xcolor}
\usepackage{stackrel}
\usepackage{accents}
\usepackage{empheq}
\usepackage[title]{appendix}

\usepackage{float}
\usepackage{braket}
\usepackage{bbold}
\usepackage{mathtools}

\DeclareMathOperator{\sgn}{sgn}

\newcommand{\numb}{\addtocounter{equation}{1}\tag{\theequation}}

\DeclareMathOperator{\Tr}{Tr}

\DeclareMathOperator{\Li}{Li}

\newcommand{\vex}[1]{\bm{\mathrm{#1}}}
\newcommand{\msf}[1]{\mathsf{#1}}

\newcommand{\dd}{\mathcal{D}}

\newcommand{\T}{\mathsf{T}}
\newcommand{\intl}[1]{\int\limits_{#1}}

\newcommand{\kb}{\vex{k}}

\newcommand{\qb}{\vex{q}}

\newcommand{\rb}{\vex{r}}
\newcommand{\xb}{\vex{r}}
\newcommand{\ww}{\omega}

\newcommand{\e}{\varepsilon}

\newcommand{\kf}{k_{\msf{F}}}
\newcommand{\vf}{v_{\msf{F}}}

\newcommand{\Tf}{E_{\msf{F}}}
\newcommand{\Ef}{E_{\msf{F}}}
\newcommand{\ktf}{k_{\msf{TF}}}

\newcommand{\phicl}{\phi_{\mathsf{cl}}}
\newcommand{\phiq}{\phi_{\mathsf{q}}}
\newcommand{\mf}{\hat{M}_F}
\newcommand{\uk}{\hat{U}_{\msf{K}}}
\newcommand{\htau}{\hat{\tau}}

\newcommand{\im}{\operatorname{Im}}
\newcommand{\re}{\operatorname{Re}}

\begin{document}
	
\title{The two-dimensional electron self-energy: Long-range Coulomb interaction}

\author{Yunxiang Liao}
\author{Donovan Buterakos}
\author{Mike Schecter}
\author{Sankar Das Sarma}%
\affiliation{Condensed Matter Theory Center and Joint Quantum Institute, Department of Physics, University of Maryland, College Park, MD 20742, USA}

\begin{abstract}
	The electron self-energy for long-range Coulomb interactions plays a crucial role in understanding the many-body physics of interacting electron systems (e.g. in metals and semiconductors), and has been studied extensively for decades.
	In fact, it is among the oldest and the most-investigated many body problems in physics.
	However, there is a lack of an analytical expression for the self-energy $\re \Sigma^{(R)}(\e,T)$ when energy $\e$ and temperature $k_{\msf{B}} T$ are arbitrary with respect to each other (while both being still small compared with the Fermi energy).
	We revisit this problem and calculate analytically the self-energy on the mass shell for a two-dimensional electron system with Coulomb interactions in the high density limit $r_s \ll 1$, for temperature $ r_s^{3/2} \ll k_{\msf{B}} T/ \Ef \ll r_s$ and energy $r_s^{3/2} \ll |\e|/\Ef \ll r_s$.
	We provide the exact high-density analytical expressions for the real and imaginary parts of the electron self-energy with arbitrary value of $\e/k_{\msf{B}} T$, to the leading order in the dimensionless Coulomb coupling constant $r_s$, and to several higher than leading orders in $k_{\msf{B}} T/r_s \Ef$  and $\e/r_s \Ef$. We also obtain the asymptotic behavior of the self-energy in the regimes $|\e| \ll k_{\msf{B}} T$ and $|\e| \gg k_{\msf{B}} T$.
	The higher-order terms have subtle and highly non-trivial compound logarithmic contributions from both $\e$ and $T$, explaining why they have never before been calculated in spite of the importance of the subject matter.
\end{abstract}	
\date{\today}

\maketitle


\section{Introduction}

In Landau's Fermi liquid theory,
an interacting Fermi system, at low excitation energies and temperatures,  is described by long-lived excitations called ``quasiparticles'' which evolve adiabatically from the corresponding excitations of the noninteracting Fermi gas as the interactions are turned on~\cite{AGD}.
The quasiparticle is well defined only when the damping of the single particle state is small, which happens at low temperatures close to the Fermi surface. In other words, the imaginary part of the retarded self-energy $\im\Sigma^{(R)}(\kb,\e)$ should be much smaller compared with $\e+\re \Sigma^{(R)}(\kb,\e)$ at low energy $\e$ in order to have well-defined quasiparticles, satisfying the Landau Fermi liquid paradigm.
The electron self-energy is a crucial quantity, which determines not only the lifetime of the quasiparticles~\cite{Quinn1958,Chaplik,Quinn1982,Zheng,Menashe,Vignale,Fujimoto,Hodges,Li2013,Malozovsky,Jalabert,Hu,dephasing,heterostructure,Jungwirt}, 
but also their effective mass~\cite{Galitski2004,DS,Zhang,Ting,Vinter,Schulze,Rice,Chubukov2003,Chubukov2004}, the renormalization factor, and many other single particle properties~\cite{Gell-Mann,Li2011,Setiawan,Galitski2005}.
It is well-established that for low $T$ ($\ll T_{\msf{F}}$) and $|\e|$ ($\ll \Ef$), $\im \Sigma^{(R)} (\varepsilon)$ goes as $T^2$ and $\varepsilon^2$ (up to logarithmic corrections) in three-dimensional (3D) and two-dimensional (2D) Fermi systems, leading to the existence of well-defined 2D and 3D Landau Fermi liquids.  By contrast, in 1D interacting Fermi systems, quasiparticles do not exist, and the one to one correspondence with the Fermi gas picture is destroyed even for infinitesimal interactions.

The calculation of the self-energy of an interacting electron system is a condensed matter problem that has been extensively studied for decades~\cite{AGD,Mahan,Chubukov2003,Chubukov2004,Chubukov2012,Hu,Galitskii1958,Engelbrecht}. 
In fact, this is among the oldest many body problems in physics, dating back to the 1950s, when field theoretic Feynman diagram techniques were first used in calculating properties of simple metals within the 3D interacting electron liquid model~\cite{AGD,Abrikosov}. Later, similar many body techniques were used to study the properties of 2D interacting electron liquids in various artificial semiconductor structures~\cite{2DES-Rev}. Most of these calculations, where the inter-electron interaction is the long-range Coulomb coupling, are either completely numerical, dubbed ``$GW$'' approximation~\cite{Hedin}, or just leading order theories in $\e$ or $T$. 
To the best of our knowledge, the expression for the self-energy $\Sigma^{(R)}(\kb,\e)$ with arbitrary $\e/k_{\msf{B}}T$ is unknown for such an interacting system with Coulomb interactions.
For this reason, we revisit the problem and calculate analytically the on-shell self-energy using the random phase approximation (RPA).
In the leading order $r_s$ expansion, where $r_s$ is the standard dimensionless Coulomb coupling parameter, we obtain the real and imaginary parts of the self-energy up to the next to the leading order $\left( \min(|\e|,T) / \Ef r_s \right)^3$. We also extract from these expressions the asymptotic behavior of $\Sigma^{(R)}(\e,T)$ in the low energy $|\e| \ll k_{\msf{B}} T$ and low temperature $ k_{\msf{B}} T \ll |\e|$ limits, with the leading order terms consistent with previous studies~\cite{Zheng}.
The higher-order generalization of the analytical self-energy expressions for the 2D electron liquid is the main result of the current work.

For the long-range Coulomb coupling, it is well-known that an asymptotically exact many-body description for the interacting self-energy is available in the high-density limit, $r_s \ll 1$, where only the infinite series of polarization diagrams (see Fig.~\ref{fig:p1}) involving the electron-hole ``bubbles'' (or ``rings'') need to be kept in the theory. This bubble diagram description of the system is equivalent (see Fig.~\ref{fig:p1}) to a theory involving the leading-order self-energy calculation in the dynamically screened Coulomb coupling (instead of the bare Coulomb coupling appearing in the Hamiltonian), where the dynamical screening is approximated by RPA.  We therefore refer to the self-energy theory in this leading order dynamical screening approximation itself as the RPA self-energy.  Such an RPA self-energy is exact to the leading-order in $r_s$, and is extensively used in materials physics, where, for historical reasons, it is universally called the ``$GW$'' approximation.  Our goal is to analytically calculate the interacting 2D self-energy to the leading-order in $r_s$ (where it is exact for the interacting problem), and to nonleading orders in $\varepsilon$ and $T$.

We emphasize that what we call RPA here is sometimes referred to as the ``$GW$ approximation" in numerical band structure theories, where $W$ refers to the dynamically RPA-screened Coulomb interaction (see Fig. ~\ref{fig:p1} below) and $G$ refers to the Green's function-- i.e. our Fig.~\ref{fig:p1}a is precisely the $GW$ approximation of band structure theories within the leading-order in $r_s$ approximation of interest in the current work.  We prefer the terminology RPA because it is the original terminology for the analytical theory and also because RPA manifestly emphasizes that the approximation involves keeping only the ring or bubble diagrams (Fig. \ref{fig:p1}b) in the screened interaction (``$W$" in band structure computations).  The theory involves neglect of vertex corrections. Indeed electron self-energy has never been calculated analytically keeping vertex corrections in the theory and it is unknown how to do it correctly and consistently.  But the issue of vertex correction is moot since our goal is to calculate the analytical self-energy in the leading order in the coupling constant $r_s$, which is exactly given just by RPA.  So, for obtaining the correct terms up to leading order in $r_s$, vertex corrections can be neglected as we do.  The same consideration applies in terms of whether one should use the full Green's function $G$ self-consistently in Fig.~\ref{fig:p1} or just the noninteracting Green's function $G_0$ in calculating the diagrams of Fig.~\ref{fig:p1}-- up to $O(r_s)$ these two approximations give the same result.  
The $GW$ approximation, which uses the full Green's function $G$ instead of the noninteracting $G_0$ done in our theory,  is perturbatively mixing orders since, in each order, vertex diagrams are left out.
Thus, the $GW$-approximation, in spite of its widespread numerical use in band structure theories, is not a consistent many-body perturbative approximation.
In addition, using the self-consistent  Green's function $G$, instead of noninteracting $G_0$, does not change the leading-order $r_s$ result.  
Up to the leading-order in $r_s$, the two theories are identical, only to higher orders in $r_s$, these two approximations, $GW$ and $G_0W$, differ.
The current theory, an expansion in ring or bubble diagrams, is the appropriate theory for an expansion in the coupling constant $r_s$. 
Our interest is obtaining the exact analytical leading order $r_s$ result, and therefore RPA and $GW$ are the same theory for us, although in detailed numerical simulations, where higher-order terms are mixed in, the two theories would differ.
Our results are exact up to the leading order in $r_s$, no approximation whatsoever-- we have kept all the diagrams necessary for the leading order in $r_s$ theory for long-range Coulomb coupling.
It has been known for a long time that the ring or bubble diagrams give the correct leading-order $r_s$ results~\cite{Gell-Mann-Brueckner}.

The rest of this paper is organized as follows. In Sec.~\ref{sec:formula}, we present the general formulas for the self-energy of a 2D electron system with Coulomb interactions. Using these formulas, in Sec.~\ref{sec:results}, we derive analytically the expressions for the imaginary and real parts of the electron self-energy in the high density ($r_s \ll 1$), low temperature ($k_{\msf{B}}T/\Ef \ll r_s$) and low energy ($|\e|/\Ef \ll r_s$) limit for arbitrary value of $\e/k_{\msf{B}}T$. The asymptotic expressions for $|\e|/ k_{\msf{B}}T \ll 1$ and $|\e|/k_{\msf{B}}T \gg 1$ are also presented in this section. In Sec.~\ref{sec:conclusion}, we conclude with a brief discussion. 
Finally, appendices are devoted to some technical details: 
In Appendix~\ref{sec:AppD}, the self-energy formulas shown in Sec.~\ref{sec:formula} are derived using the Keldysh technique. We provide in Appendix~\ref{sec:momentum} the evaluation of the momentum integrals appearing in the self-energy formulas, and in Appendix~\ref{sec:integrals} the frequency integrals involving hyperbolic $\tanh$ and $\coth$ functions.

\section{General formulas}\label{sec:formula}

\begin{figure}[t!]
	\centering
	\includegraphics[width=0.95\linewidth]{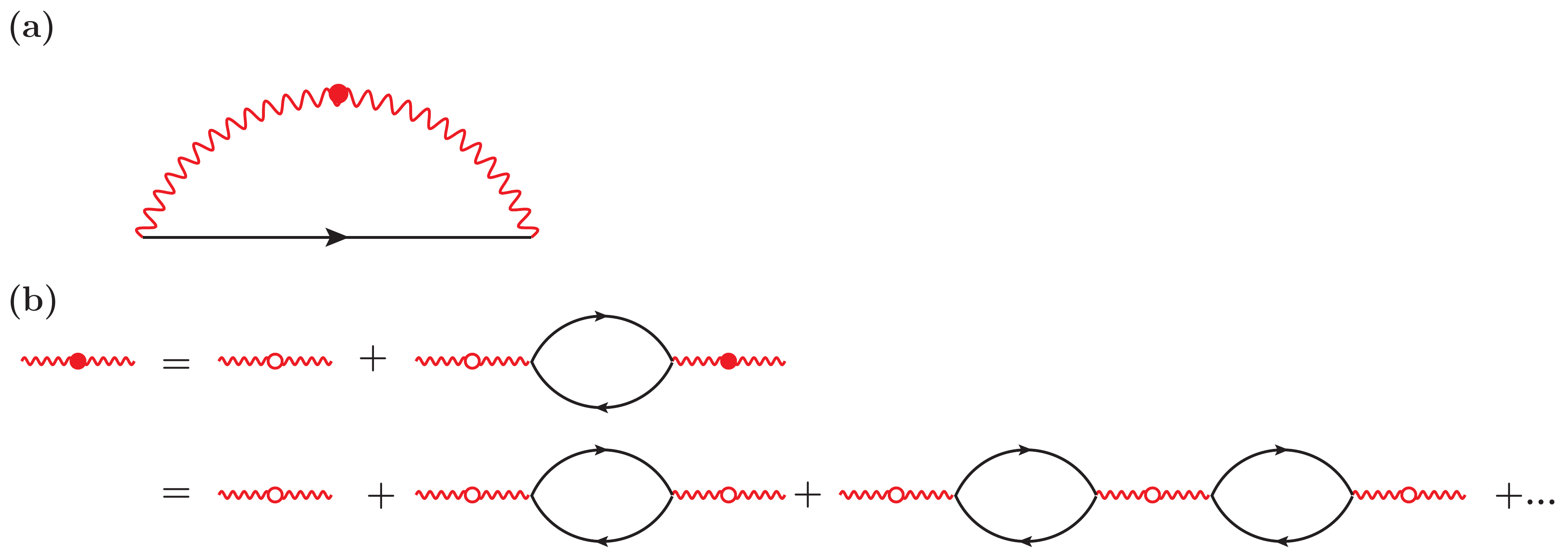}
	\caption{
		(a) The RPA self-energy diagram for a system of interacting electrons.
		(b) The diagrammatic definition of the RPA dynamically screened interaction. 
		In both panels, the black line represents the bare electron Green's function, while the red wavy line with a solid (open) dot corresponds to the RPA dynamically screened interaction (bare interaction). 
		The RPA interaction is the dressed Green's function for the Hubbard-Stratonovich field that decouples the interactions, and is given by the infinite Dyson series with repeated insertion of the polarization bubbles, as shown in panel (b).
	}
	\label{fig:p1}
\end{figure}

This section summarizes the general formulas for the electron self-energy. 
The results provided in this section are, in principle, known, but we give them here for the sake of completeness because we have not seen them written down anywhere in the literature  in the precise form necessary for our calculations.  In addition, this section provides a context and serves the useful purpose of explaining our notations and the actual calculations as well as the analytical results.
We consider a clean 2D electron system with Coulomb interactions, a parabolic energy dispersion and a spin degeneracy factor of 2,
and from now on adopt the units $k_{\msf{B}}=\hbar=1$. 
(Spin is an implicit variable since we only consider a paramagnetic situation with no explicit spin-dependent scattering-- the only interaction in the problem is the long-range Coulomb coupling, which is spin-independent.)
The detailed derivation of these formulas is presented in Appendix~\ref{sec:AppD},
while an alternative Matsubara approach can be found in Ref.~\cite{AGD}.

To the lowest order in the dynamically screened interaction, the retarded electron self-energy is given by
\begin{align}\label{eq:Sig1}
\begin{aligned}
	&\Sigma^{(R)} (\kb,\e)
	=\,
	\frac{i}{2}
	\intl{\qb,\ww}
	\left\lbrace 
	D^{(K)}(\qb,\ww)G_0^{(R)}(\kb+\qb,\e+\ww)
	\right. 
	\\
	&\left. 
	\qquad
	+
	D^{(A)}(\qb,\ww)
	G_0^{(K)}(\kb+\qb,\e+\ww)
	\right\rbrace,
\end{aligned}
\end{align}
and the corresponding self-energy diagram in plotted in Fig.~\ref{fig:p1}(a).
Here we have employed the shorthand notation $\int_{\qb}\equiv \int d^2q/(2\pi)^2$ and 
$\int_{\ww}\equiv \int_{-\infty}^{\infty} d \ww /2\pi$.

$G_0$ denotes the non-interacting fermionic Green's function, and is represented diagrammatically by the black line in Fig.~\ref{fig:p1}.
Its retarded (advanced) component acquires the form
\begin{align}	\label{eq:GR}
		G_0^{(R)/(A)}(\kb,\e)	= \left[\e - \xi_{\kb} \pm i \eta  \right]^{-1}.
\end{align}		
$\eta$ is a positive infinitesimal, and $\xi_{\kb}\equiv k^2/2m-\mu$, with $\mu \equiv \Ef$, where $\Ef$ is the noninteracting Fermi energy,  being the chemical potential. 
We use $\kf$ for the Fermi momentum, defined by $\Ef=\kf^2/2m$.  We note that the dimensionless Coulomb coupling (or the effective fine structure constant) is simply given by $r_s=\sqrt{2} e^2/\vf$, where the Fermi velocity $\vf=\kf/m$.
The Keldysh Green's function is related to its retarded and advanced counterparts through the fluctuation-dissipation theorem (FDT):		
\begin{align}	
		\label{eq:FDT-G}	
		&
		G_0^{(K)}(\kb,\e)=\left[ G_0^{(R)}(\kb,\e)-G_0^{(A)}(\kb,\e)\right] \tanh(\e/2T),
\end{align}	
and, unlike the other two components, depends on the occupation number.
Hereafter, superscripts $(R)$, $(A)$ and $(K)$ stand for the retarded, advanced, and Keldysh components, respectively.

In Eq.~\ref{eq:Sig1}, $D$ indicates the RPA dynamically screened interaction,
which is represented by the red wavy line with a solid dot in Fig.~\ref{fig:p1}.
It can be considered as the dressed Green's function for the bosonic field that, through the Hubbard-Stratonovich (H.S.) transformation, decouples the interactions (see Appendix~\ref{sec:AppD} for details).
In Fig.~\ref{fig:p1}(b), the RPA interaction $D$ is defined diagrammatically by the infinite Dyson series of the polarization bubble diagrams where the red wavy line with an open dot stands for the 2D bare Coulomb potential $V(q)=2\pi e^2/q$, and the black bubble corresponds to the polarization operator whose retarded component is given by
\begin{align}\label{eq:Pi2}
\begin{aligned}
	&\Pi^{(R)} (\qb,\ww)
	=\,
	-i
	\intl{\kb,\e}
	\left[ 
	G_0^{(R)}(\kb+\qb,\e+\ww)G_0^{(K)}(\kb,\e)
	\right.
	\\
	&
	\left. 
	\qquad +
	G_0^{(K)}(\kb+\qb,\e+\ww)G_0^{(A)}(\kb,\e)
	\right].
\end{aligned}
\end{align}
It is therefore straightforward to see that the retarded RPA dynamically screened interaction can be extracted from the following Dyson equation
\begin{align}\label{eq:D1-a}
\begin{aligned}
	D^{(R)}(\qb,\ww)
	=\,&
	\left[ V^{-1}(q)-\Pi^{(R)}(\qb,\ww)\right]^{-1},
\end{aligned}
\end{align}
while its advanced and Keldysh components are related to the retarded one through
\begin{subequations}\label{eq:Da}
\begin{align}\label{eq:D1-b}
	&\begin{aligned}
	D^{(A)}(\qb,\ww)
	=&
	\left[ D^{(R)}(\qb,\ww)\right]^*,
	\end{aligned}
	\\
	&\begin{aligned}\label{eq:FDT-D}
	D^{(K)}(\qb,\ww)
	=\,&
	\left[ D^{(R)}(\qb,\ww)-D^{(A)}(\qb,\ww)\right] \coth\left(\ww/2T \right),
	\end{aligned}
\end{align}
\end{subequations}
as expected for a bosonic propagator.
Here the last equation constitutes the FDT relation between the components of the RPA interaction $D$.

In the static limit $\ww \ll \kf q/m$, 
 $-\Pi^{(R)}(\qb,\ww)$ is well approximated by $\nu=m/\pi$, the density of states at the Fermi level. As a result,  $D^{(R)/(A)}(\qb,\ww)$ is reduced to the static screened interaction
\begin{align}
\begin{aligned}
\tilde{V}(q)
\equiv 
\frac{1}{V^{-1}(q)+\nu}
=
\dfrac{1}{\nu} \dfrac{\ktf}{q+\ktf},
\end{aligned}
\end{align}
with $\ktf \equiv 2\pi e^2 \nu$ being the Thomas-Fermi screening wavevector.	

Inserting the explicit expression for the non-interacting electron Green's function $G_0$ (Eq.~\ref{eq:GR}) into Eq.~\ref{eq:Sig1}, and utilizing the FDT relation as well as the Kramers-Kr\"{o}nig relation (Eq.~\ref{eq:KK})
for both $G_0$ and $D$, one finds that the imaginary and real parts of the electron self-energy are given by
\begin{widetext}
\begin{subequations}\label{eq:Sig}
	\begin{align}
	&\begin{aligned}\label{eq:ImSig-0}
	\im \Sigma^{(R)} (\kb,\e)
	=\,
	\frac{m}{4 \pi^2k}
	\int_{-\infty}^{\infty} d\ww
	\left[ \coth\left( \frac{\ww}{2T}\right) -\tanh\left( \frac{\ww+\e}{2T}\right) \right]
	\int_{q_-(\ww)}^{q_+(\ww)}  d q
	\dfrac{ \im D^{(R)}(\qb,\ww)}{
	\sqrt{1 -\left[ \frac{m}{kq}\left( \ww + \Delta \e \right)\right] ^2   }  },
	\end{aligned}
	\\		
	&\begin{aligned}\label{eq:ReSig-0}
	\re \Sigma^{(R)} (\kb,\e)
	=\,&
	\frac{m}{4 \pi^2 k}
	\int_{-\infty}^{\infty} d\ww
	\tanh\left( \frac{\e+\ww}{2T}\right)
	\int_{q_-(\ww)}^{q_+(\ww)} d q
	\dfrac{	\re D^{(R)}(\qb,\ww) }{
		\sqrt{1-\left[  \frac{m }{kq}\left( \ww +\Delta \e\right) \right] ^2   } 		
	}
	\\
	-&
	\frac{m}{4 \pi^2 k}
	\int_{-\infty}^{\infty} d\ww
	\coth\left( \frac{\ww}{2T}\right)
	\left( \int_0^{q_-(\ww)} d q +\int_{q_+(\ww)}^{\infty} d q \right) 
	\im D^{(R)}(\qb,\ww)
	\frac{	\sgn \left(\ww+  \Delta \e \right) }{\sqrt{\left[ \frac{m}{kq}\left( \ww+\Delta \e \right)\right] ^2-1}}.
	\end{aligned}
	\end{align}
\end{subequations}	
\end{widetext}
For simplicity, here we have defined
\begin{subequations}
	\begin{align}
	&\begin{aligned}\label{eq:de}
	\Delta \e  \equiv \e-\xi_{\kb}-q^2/2m ,
	\end{aligned}
	\\
	&
	\begin{aligned}\label{eq:q}
	q_{\pm}(\ww)	\equiv \left|  \pm  k + \sqrt{k^2 + 2m\left( \ww+\e -\xi_{\kb}\right) } \right|.
	\end{aligned}
		\end{align}
\end{subequations}
	
Equations ~\ref{eq:ImSig-0} and~\ref{eq:ReSig-0} can be directly numerically calculated for arbitrary $r_s$, $\varepsilon$, and $T$ to provide the self-energy function of a 2D interacting electron liquid.  Our goal is to obtain the analytical expressions for small values of $r_s$, $\varepsilon$, and $T$ as described in the next section.  We note that formally Eqs.~\ref{eq:ImSig-0} and~\ref{eq:ReSig-0} appear to be 2D integrals over $\omega$ and $q$, but this is misleading since the screened interaction $D$ itself (see Eqs.~\ref{eq:Pi2}-\ref{eq:Da} above) is formally a 3D integral.  Thus, Eq.~\ref{eq:Sig} in general defines a highly singular five-dimensional integral, which is not easy to handle directly numerically although its $T=0$ version has been calculated numerically~\cite{Jalabert} and crude numerical calculations have also been performed for the temperature dependent 2D self-energy by approximating the dynamically screened interaction $D$ as a simple function with poles within the so-called plasmon pole approximation~\cite{DS1979}. There had also been an early purely numerical attempt to calculate the 2D self-energy at very high $(T \gg \Ef)$ temperatures, where the static screening approximation was used to replace the dynamically screened interaction~\cite{Nakamura}. Such numerical self-energy calculations carried out with simplistic and uncontrolled approximations fail to provide any analytical insight into the low-temperature quasiparticle properties of the interacting 2D Fermi liquid, which is the goal of our study.	
	
\section{Results}\label{sec:results}

In the previous section, the real and imaginary parts of the electron self-energy $\Sigma^{(R)}(\kb,\e)$ are expressed as two-variable integrals in terms of the retarded RPA interaction $D^{(R)}(\qb,\ww)$ given by Eq.~\ref{eq:D1-a}. 
Having these formulas, we now calculate analytically the on-shell ($\e=\xi_{\kb}$) self-energy $\Sigma^{(R)}(\kb,\e)$ close to the Fermi surface ($k \approx \kf$) in the high density limit ($r_s \ll 1$), where the RPA ring diagram approximation should be exact for the Coulomb coupling. We work in the regime where $r_s^{3/2} \Tf \ll \Delta  \ll r_s \Tf $, with $\Delta= \left\lbrace T, |\e| \right\rbrace $, and obtain the electron self-energy up to the order of $\left( \Delta / \Ef r_s \right)^3$ and the leading order in $r_s$.

In the low temperature regime, the polarization operator $\Pi^{(R)}(\qb,\ww)$ can be approximated by its zero temperature result, whose explicit form has been calculated in Ref.~\cite{Stern} and may be expressed in terms of two dimensionless variables: $q/\kf$ and $m\ww/\kf q$ (see Eq.~\ref{eq:Pi0}).
Because of the RPA interaction and the thermal factors, the most significant contribution to the integrals in Eq.~\ref{eq:Sig} comes from the region
$ q \lesssim \ktf =\sqrt{2}r_s \kf $
and 
$\ww \lesssim \max (T, |\e|)$.
As a result, terms of higher orders in $q/\kf$ and $m\ww/\kf q$ in the integrands lead to, respectively, higher order terms in $r_s$ and $\max (T,|\e|)/\Ef r_s$ in the electron self-energy $\Sigma^{(R)}(\kb,\e)$, and are negligible in the high density ($r_s \ll 1$), low temperature ($T/\Tf \ll r_s$) and small energy ($|\e|/\Tf \ll r_s$)  limit.
It is therefore only necessary to keep the first few leading terms in $q/\kf$ and $m\ww/\kf q$ in the integrals to obtain the leading results of interest to us.

\subsection{Imaginary part of the self-energy}

To calculate the imaginary part of the self-energy on the mass shell ($\e=\xi_{\kb}$), we insert the expression for the polarization operator~\cite{Stern} into Eq.~\ref{eq:ImSig-0} and set $\Delta \e = -q^2/2m$.
After the momentum integration, this yields
\begin{widetext}
\begin{subequations}\label{eq:ImS-0I}
\begin{align}
	&\begin{aligned}\label{eq:ImS-0}
		\im \Sigma^{(R)} (\e)
		=\,&
		\int_{0}^{\infty} \frac{d\ww}{2\pi}
		\left[2 \coth\left( \frac{\ww}{2T}\right) 
		-\tanh\left( \frac{\ww+\e}{2T}\right)-\tanh\left( \frac{\ww-\e}{2T}\right)
		\right]
		\im  I(\ww),
	\end{aligned}	
		\\
	&\begin{aligned}\label{eq:I_1}
		\im I(\ww)
		=\, &
		-
		\left\lbrace 
		\frac{1}{4}\frac{|\ww|}{\Ef}
	    \left[ \ln \left( \frac{2\sqrt{2} r_s \Ef}{|\ww|}\right) -1 \right] 
	    +\frac{1}{2\sqrt{2}r_s} \left( \frac{\ww}{\Ef}\right)^2 \right\rbrace 
	    \sgn \ww.
	\end{aligned}
\end{align}
\end{subequations}
\end{widetext}
Here $I(\ww)$ is defined as the integral
\begin{align}\label{eq:I}
	I(\ww)
	\equiv
	\frac{m}{2 \pi k}
	\int_{q_-(\ww)}^{q_+(\ww)} d q
	\dfrac{	D^{(R)}(\vex{q},\ww) }{
		\sqrt{1-\left[  \frac{m }{kq}\left( \ww -\frac{q^2}{2m} \right) \right] ^2   } 		
	}.
\end{align}
Its detailed calculation is shown in Appendix~\ref{sec:momentum}.

The frequency integration in Eq.~\ref{eq:ImS-0} can be done by expressing the hyperbolic functions as infinite exponential series:
\begin{align}\label{eq:expser}
\begin{aligned}
	&
	\tanh (x)
	=\,
	1+2\sum_{k=1}^{\infty} (-1)^k e^{-2kx},
	\qquad
	&&x>0,
	\\
	&
	\coth (x)
	=\,
	1+2\sum_{k=1}^{\infty}  e^{-2kx},
	\qquad
	&&x>0.
\end{aligned}
\end{align}
We evaluate integrals of such forms in Appendix~\ref{sec:integrals},
and use the results (Eq.~\ref{eq:hy-Im}) to obtain the imaginary part of the self-energy on the mass shell
\begin{widetext}
\begin{align}\label{eq:ImSig}
\begin{aligned}
	\im \Sigma^{(R)}  ( \e)
	=&
	-\frac{T^2}{\Tf}  \ln \left( \frac{\sqrt{2} r_s \Tf}{T}\right) 
	g_1(\frac{\e}{T})
	-
	\frac{T^2}{\Tf} 
	g_2(\frac{\e}{T})
	-
	\frac{T^3}{r_s \Tf^2}
	g_3(\frac{\e}{T}),
\end{aligned}
\end{align}
\end{widetext}
where
\begin{align}\label{eq:g12}
\begin{aligned}
	g_1(\frac{\e}{T})
	\equiv\,&
	\frac{1}{8 \pi }\left( \pi^2+\frac{\e^2}{T^2}\right),
	\\
	g_2(\frac{\e}{T})
	\equiv \,&
	-\frac{\pi}{24}
	\left(6- \gamma_E -\ln \frac{2}{\pi^2} -24\ln A \right) 
	\\
	&
	-\frac{\left(2-\gamma_E-\ln 2 \right)}{8 \pi } \frac{\e^2}{T^2} 
	\\
	&
	+\frac{1}{4 \pi }
	\left[ \partial_s \Li_s (-e^{-\e/T})+ \partial_s \Li_s (-e^{\e/T})\right] \bigg \lvert_{s=2} ,
	\\
	g_3(\frac{\e}{T})
	\equiv \,&
	\frac{\sqrt{2}}{\pi}
	\left[ 
	\zeta(3)
	-\frac{1}{2}\Li_3 (-e^{\e/T})	
	-\frac{1}{2}\Li_3 (-e^{-\e/T})
	\right]. 
\end{aligned}
\end{align}
Here 
$\Li_s(z)=\sum_{k=1}^{\infty} z^k/k^s$ denotes the polylogarithm function and $\zeta(s)=\Li_s(1)$ represents the Riemann zeta function.
$\gamma_E \approx 0.577216$ is the Euler's constant and $A \approx 1.28243$ is the Glaisher's constant.  
Following a straightforward calculation, we find from Eq.~\ref{eq:ImSig} the asymptotic behavior of $\im \Sigma^{(R)}(\e)$ in the regime $|\e| \ll T$:
\begin{align}\label{eq:ImSig-<}
\begin{aligned}
	&\im \Sigma^{(R)} (|\e |\ll T)
	=\,
	-\frac{\pi}{8} 
	\frac{ T^2 }{\Tf} 
	\ln \left( \frac{\sqrt{2} r_s \Tf}{T}\right) 
	\\
	&
	+
	 \frac{\pi}{24}
	\left(6 + \ln 2\pi^3 -36 \ln A \right) 
	\frac{ T^2}{\Tf}
	- \frac{7\zeta(3)}{2\sqrt{2}\pi} \frac{T^3}{r_s\Tf^2}.
\end{aligned}
\end{align}

\begin{figure}[t!]
	\centering
	\includegraphics[width=0.95\linewidth]{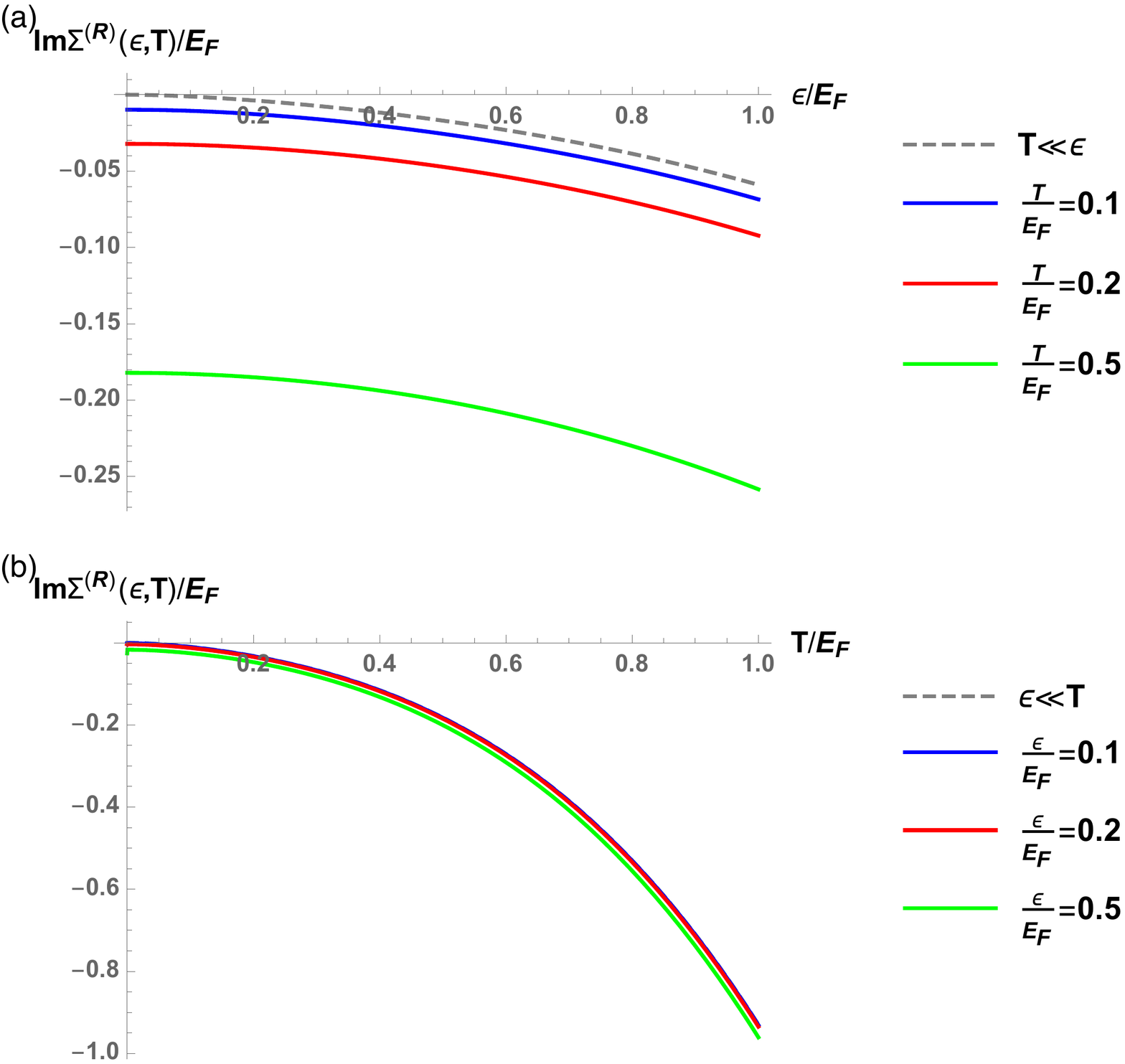}
	\caption{
		Imaginary part of the on-shell electron self-energy $\im \Sigma^{(R)} (\e,T)$.
		In panel (a) [(b)], $\im \Sigma^{(R)} (\e,T)/\Tf$ given by Eq.~\ref{eq:ImSig} is plotted as a function of $\e/\Tf$ ($T/\Ef$) for different values of $T/\Ef$ ($\e/\Ef$) and is compared with the asymptotic result for $T \ll \e$ ($\e \ll T$) given by Eq.~\ref{eq:ImSig->} (Eq.~\ref{eq:ImSig-<}). Solid curves from top to bottom correspond to $\im \Sigma^{(R)} (\e,T)/\Tf$ at $T/\Ef$ ($\e/\Ef$) equals to $0.1$, $0.2$ and $0.5$ , while dashed curve represents the the asymptotic result for $T \ll \e$ ($\e \ll \T$).
		In both panels, $r_s$ is set to be $1$.
	}
	\label{fig:p2}
\end{figure}

Eq.~\ref{eq:ImSig} can be rewritten as
\begin{align}
\begin{aligned}
	\im \Sigma^{(R)} ( \e)=&
	-
	\frac{\e^2}{ \Tf}   \ln \left( \frac{\sqrt{2} r_s \Tf}{ |\e| }\right) 
	\tilde{g}_1(\frac{\e}{T})
	-
	\frac{\e^2}{\Tf} 
	\tilde{g}_2(\frac{\e}{T})
	\\
	&-
	\frac{|\e|^3}{r_s \Tf^2}
	\tilde{g}_3(\frac{\e}{T}),
\end{aligned}
\end{align}
where we have defined 
\allowdisplaybreaks
\begin{align*}
	&\tilde{g}_1(\frac{\e}{T})
	\equiv \,
	\frac{T^2}{\e^2} g_1(\frac{\e}{T}),
	\\
	&
	\tilde{g}_2(\frac{\e}{T})
	\equiv\,
	\ln\left( \frac{|\e|}{T}\right)
	\frac{T^2}{\e^2} 
	g_1(\frac{\e}{T})
	+ \frac{T^2}{\e^2}
	g_2(\frac{\e}{T}),
	\\
	&
	\tilde{g}_3(\frac{\e}{T})
	\equiv \,
	\frac{T^3}{|\e|^3} g_3(\frac{\e}{T}).
	\numb
\end{align*}
This leads to the following asymptotic expression for $\im \Sigma^{(R)} (\e)$ when $|\e| \gg T$:
\begin{align}\label{eq:ImSig->}
\begin{aligned}
	&\im \Sigma^{(R)} (|\e| \gg T)
	=\,
	-\frac{\e^2 }{8 \pi \Tf}  \ln \left( \frac{\sqrt{2} r_s \Tf}{|\e| }\right) 
	\\
	&
	-
	\frac{\e^2 }{8 \pi \Tf} 
	 \left( \ln 2 -\frac{1}{2} \right)
	 -
	 \frac{1}{6\sqrt{2} \pi}\frac{|\e|^3}{r_s \Tf^2}.
\end{aligned}
\end{align}
We note that the leading order terms (i.e. the first terms on the right hand sides) in the asymptotic expressions Eqs.~\ref{eq:ImSig-<} and~\ref{eq:ImSig->} are consistent with the results in Ref.~\cite{Zheng}, which obtained the correct leading order $T^2 \ln T$  and $\varepsilon^2 \ln \varepsilon$ forms for the 2D imaginary self-energy.  We note that in spite of the additional logarithmic factors compared with the corresponding 3D self-energy~\cite{AGD,Quinn1958}, the imaginary self-energy has the quadratic dependence on $T$ and/or $\varepsilon$, indicating that the interacting 2D system is a Landau Fermi liquid at low temperatures and excitation energies.  Our results in Eqs.~\ref{eq:ImSig-<} and~\ref{eq:ImSig->} provide the full analytical form for the 2D imaginary self-energy including the next to leading order terms in the excitation energy and temperature.

In Fig.~\ref{fig:p2}(a) [Fig.~\ref{fig:p2}(b)], we plot the imaginary part of the on-shell self-energy given in Eqs.~\ref{eq:ImSig} and~\ref{eq:g12} as a function of $\e/\Ef$ ($T/\Ef$) for different values of $T/\Ef$ ($\e/\Ef$), together with the asymptotic expression for $T \ll \e$ ($\e \ll T$) given in Eq.~\ref{eq:ImSig->} [Eq.~\ref{eq:ImSig-<} ]. 
The solid curves from top to bottom are associated with $\e/\Ef$ ($T/\Ef$) equals to $0.1$, $0.2$ and $0.5$.
As the value of $\e/\Ef$ ($T/\Ef$) decreases, the corresponding solid curve approaches the dashed one, which represents the analytical asymptotic result for $T \ll \e$ ($\e \ll T$) as given in Eq.~\ref{eq:ImSig->} [Eq.~\ref{eq:ImSig-<}].

\subsection{Real part of the self-energy}

Compared with the imaginary part of the self-energy, the calculation of the real part is much more difficult since it requires one more integration involving a branch cut.
For the real part of the self-energy on the mass shell ($\e=\xi_{\kb}$), the second integral in Eq.~\ref{eq:ReSig-0} vanishes to the leading order in $r_s$, while the first integral, after the momentum integration (for details, see Appendix~\ref{sec:momentum}),
is further reduced to 
\begin{widetext}
\begin{subequations}
\begin{align}
	&\begin{aligned}\label{eq:ReS-0}
	\re \Sigma^{(R)} (\e)
	=\,
	\int_{0}^{\infty} 
	\frac{d\ww}{2\pi}
	\left[ 
	\tanh\left( \frac{\ww+\e}{2T}\right)
	-
	\tanh\left( \frac{\ww-\e}{2T}\right)
	\right] 
	\re I(\ww),
	\end{aligned}
	\\
	&\begin{aligned}\label{eq:I_2}
		\re I(\ww)
		=
		\frac{r_s}{\sqrt{2}}
		\left[ 
		\ln \left( \frac{2\sqrt{2}}{r_s} \right) 
		-\frac{\pi}{4\sqrt{2}r_s}   \frac{|\ww|}{\Tf} 
		+\frac{5 }{16 r_s^2} \frac{\ww^2}{\Tf^2} \ln \left( \frac{4 \sqrt{2} r_s \Tf}{|\ww|}\right) 
		-\frac{17}{32 r_s^2} \frac{\ww^2}{\Tf^2}
		\right]. 
	\end{aligned}
\end{align}
\end{subequations}
\end{widetext}

We then evaluate the $\ww-$integration in Eq.~\ref{eq:ReS-0} by utilizing the exponential expansion of the hyperbolic function (Eq.~\ref{eq:expser}).
The detailed calculation is relegated to Appendix~\ref{sec:integrals}.
From Eq.~\ref{eq:hy-Re}, we find that the real part of the on-shell self-energy has the following asymptotic form for low energies and temperatures
\begin{align}\label{eq:ReSig}
\begin{aligned}
	\re \Sigma^{(R)} (\e)
	=\,&
	h_0 \e
	+
	\frac{T \e}{\Tf}
	 h_1 \left( \frac{\e}{T}\right)
	+
	\frac{ T^2 \e}{r_s \Tf^2} 
	\ln \left( \frac{ r_s \Tf}{ T }\right)
	h_2 \left( \frac{\e}{T}\right) 
	\\
	&
	+
	\frac{ T^2 \e}{r_s \Tf^2} 
	h_3 \left( \frac{\e}{T}\right) 
	,
\end{aligned}
\end{align}
where
\begin{align}\label{eq:h}
\begin{aligned}
	&
	h_0 \equiv 
	\frac{r_s}{\sqrt{2} \pi}
	\ln \left( \frac{2\sqrt{2}}{r_s} \right), 
	\\
	&
	h_1 \left( \frac{\e}{T}\right) \equiv
	-
	\frac{1}{8}	
	\frac{T}{\e}
	\left[ \Li_2(-e^{-\frac{\e}{T}}) - \Li_2(-e^{\frac{\e}{T}}) \right],
	\\
	&
	h_2 \left( \frac{\e}{T}\right)  \equiv
	\frac{5}{48\sqrt{2} \pi}
	\left(  \pi^2 +  \frac{\e^2}{T^2} \right) , 
	\\
	&
	h_3 \left( \frac{\e}{T}\right)  \equiv
	-
	\frac{1}{96 \sqrt{2} \pi }
	\left( 32-10\gamma_E-25\ln 2 \right) 
	\left( \frac{\e^2}{T^2} +\pi^2  \right) 
	\\
	&-
	\frac{5}{8 \sqrt{2} \pi }
	\frac{T}{\e}
	\left[ \partial_s \Li_s (-e^{-\frac{\e}{T}}) - \partial_s \Li_s (-e^{\frac{\e}{T}}) \right] \bigg\lvert_{s=3}.
\end{aligned}
\end{align}
The subleading term  $T \varepsilon h_1(\frac{\varepsilon}{T})/\Ef $ in Eq.~\ref{eq:ReSig} is of a form similar to the result obtained for a model with short-range interaction in Refs.~\cite{Chubukov2003,Chubukov2004}. Our system is completely different, i.e., an electron system with the realistic long-range Coulomb interaction, so it is important that long-range and short-range interactions lead to similar subleading terms in the self-energy.  We emphasize that we also obtain additional terms of higher orders in $T/\Ef$ or $\varepsilon/\Ef$, which was not done in Refs.~\cite{Chubukov2003,Chubukov2004}

For $|\e| \ll T$, Eq.~\ref{eq:ReSig} becomes
\begin{align}\label{eq:ReSig-<}
\begin{aligned}
&\re \Sigma^{(R)} (|\e| \ll T)
\\
=\,&
\frac{r_s}{\sqrt{2} \pi}
\ln \left( \frac{2\sqrt{2}}{r_s} \right) 
\e
-
\frac{ \ln 2}{4} 
\frac{T \e}{\Tf}
+
\frac{5\pi}{48\sqrt{2} r_s}
\frac{ T^2 \e }{\Tf^2} 
\ln \left( \frac{r_s \Tf}{ T }\right)
\\
&
+
\frac{ T^2 \e }{\Tf^2 r_s} 
\left[ -
\frac{\pi}{96 \sqrt{2} }
\left(32-10\gamma_E-25\ln 2 \right) 
\right. 
\\
&
\left. 
\left. 
\qquad\qquad
-
\frac{5}{8 \sqrt{2} \pi }
\left( \zeta ' (2) + \frac{\pi^2}{6}\ln 2\right) 
\right] 
\right\rbrace .
\end{aligned}
\end{align}
We note that the quasiparticle effective mass obtained from this asymptotic expression is consistent with the result in Ref.~\cite{Galitski2004}.

\begin{figure}[t!]
	\centering
	\includegraphics[width=0.95\linewidth]{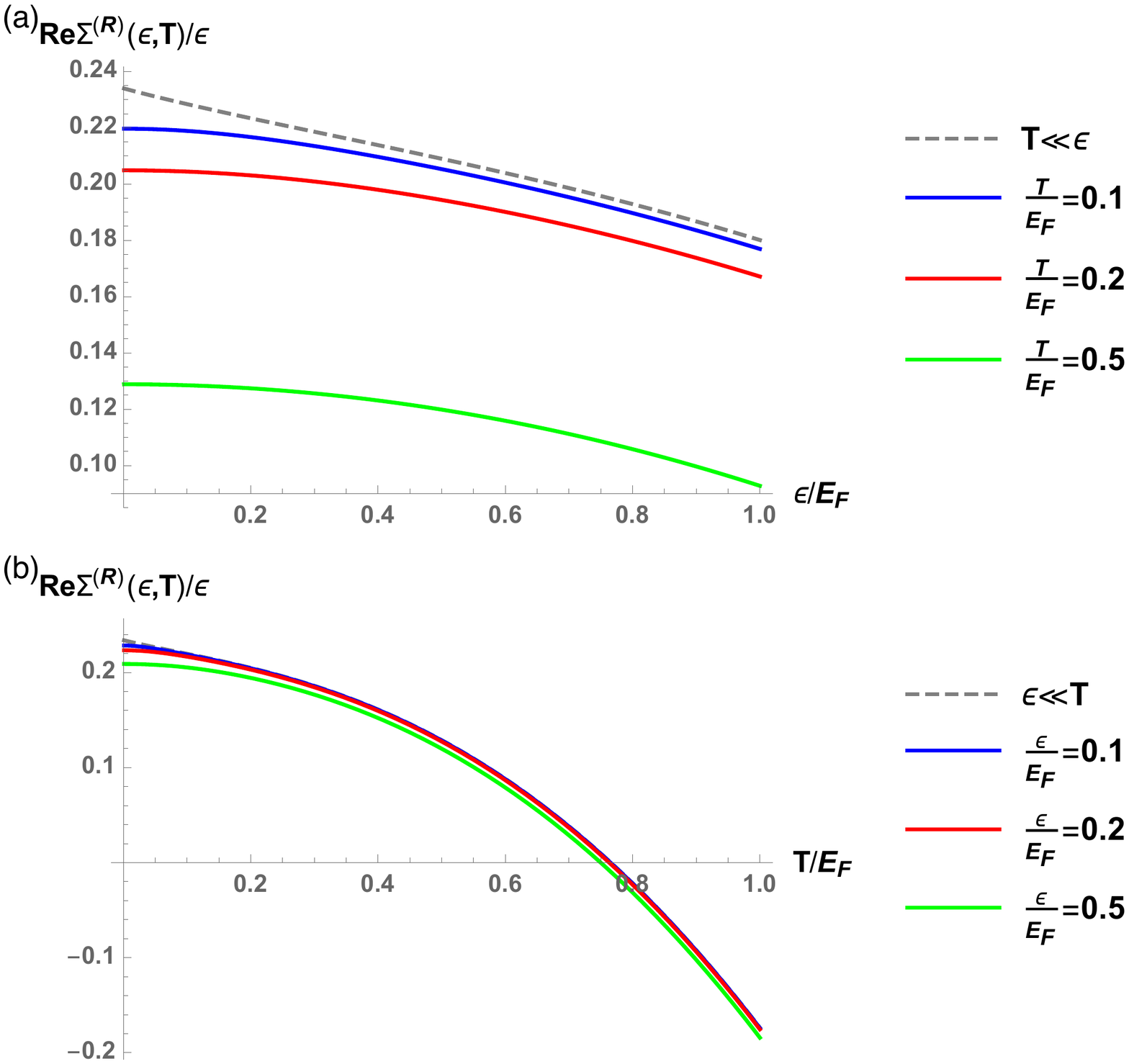}
	\caption{
		Real part of the on-shell electron self-energy $\re \Sigma^{(R)} (\e,T)$.
		In panel (a) [(b)], $\re \Sigma^{(R)} (\e,T)/\e$ given by Eq.~\ref{eq:ReSig} is plotted as a function of $\e/\Tf$ ($T/\Ef$) at various $T/\Ef$ ($\e/\Ef$) which admits the values of $0.1$, $0.2$ and $0.5$ (correspond to solid curves from top to bottom),  and is compared with the asymptotic result for $T \ll \e$ ($\e \ll T$) given by Eq.~\ref{eq:ReSig->} (Eq.~\ref{eq:ReSig-<}) represented by the dashed curve.
		$r_s=1$ in both panels.
	}
	\label{fig:p3}
\end{figure}

We then rewrite Eq.~\ref{eq:ReSig} as
\begin{align}
\begin{aligned}
&\re \Sigma^{(R)} (\e)
\\
=\,&
h_0 
\e
+
\frac{\e|\e|}{\Tf}
\tilde{h}_1 \left( \frac{\e}{T}\right)
+
\frac{ \e^3}{r_s \Tf^2} \ln \left( \frac{r_s \Tf}{| \e| }\right)
\tilde{h}_2 \left( \frac{\e}{T}\right)
\\
&+
\frac{ \e^3}{r_s\Tf^2} 
\tilde{h}_3 \left( \frac{\e}{T}\right),
\end{aligned}
\end{align}
where
\begin{align}
\begin{aligned}
&
\tilde{h}_1 \left( \frac{\e}{T}\right)
\equiv
\frac{T}{|\e|} h_1 \left( \frac{\e}{T}\right), 
\\
&
\tilde{h}_2 \left( \frac{\e}{T}\right)
\equiv
(\frac{T}{\e})^2
h_2 \left( \frac{\e}{T}\right), 
\\
&
\tilde{h}_3 \left( \frac{\e}{T}\right)
\equiv
(\frac{T}{\e})^2 \ln(\frac{|\e|}{T})
h_2 \left( \frac{\e}{T}\right)
+
(\frac{T}{\e})^2h_3 \left( \frac{\e}{T}\right).
\end{aligned}
\end{align}
From this equation above, we arrive at the asymptotic form of $\re \Sigma^{(R)} (\e)$ for $|\e| \gg T$:
\begin{align}\label{eq:ReSig->}
\begin{aligned}
	&\re \Sigma^{(R)} (|\e| \gg T)
	=\,
	\frac{r_s}{\sqrt{2} \pi}
	\ln \left( \frac{2\sqrt{2}}{r_s} \right) 
	\e
	-
	\frac{1}{16}\frac{\e|\e|}{\Tf}
	\\
	&
	+
	\frac{5 }{48\sqrt{2}\pi }\frac{ \e^3}{r_s\Tf^2} \ln \left( \frac{ r_s \Tf}{ |\e| }\right)
	+
	\frac{-41+75\ln 2}{288 \sqrt{2} \pi }\frac{ \e^3}{r_s\Tf^2}.
\end{aligned}
\end{align}

We note that, although the leading-order dependence of the real part of the self-energy on the excitation energy and temperature manifests the expected linear-in-$\varepsilon$ behavior, the next-to-leading-order terms as shown in Eqs.~\ref{eq:ReSig-<} and~\ref{eq:ReSig->} are nontrivial, and impossible to guess because of the logarithmic factors which disallow for a simple dimensional argument. 
The analytical result of the real part of the on-shell self-energy given by Eqs.~\ref{eq:ReSig} and~\ref{eq:h} is presented in Fig.~\ref{fig:p3}.
Since the leading order term in $\re \Sigma^{(R)}(\e)$ scales as $\e$, here we plot the ratio $\re \Sigma^{(R)}(\e)/\e$ instead of $\re \Sigma^{(R)}(\e)$ itself. 
In Fig.~\ref{fig:p3}a (Fig.~\ref{fig:p3}b), $\re \Sigma^{(R)}(\e)/\e$ is shown as a function of $\e/\Ef$ ($T/\Ef$) for different value of $T/\Ef$ ($\e/\Ef$), and is compared with the corresponding asymptotic expression for $T \ll \e$ ($\e \ll T$) as given in Eq.~\ref{eq:ReSig->} [ Eqs.~\ref{eq:ReSig-<}].

\begin{figure}[t!]
	\centering
	\includegraphics[width=0.95\linewidth]{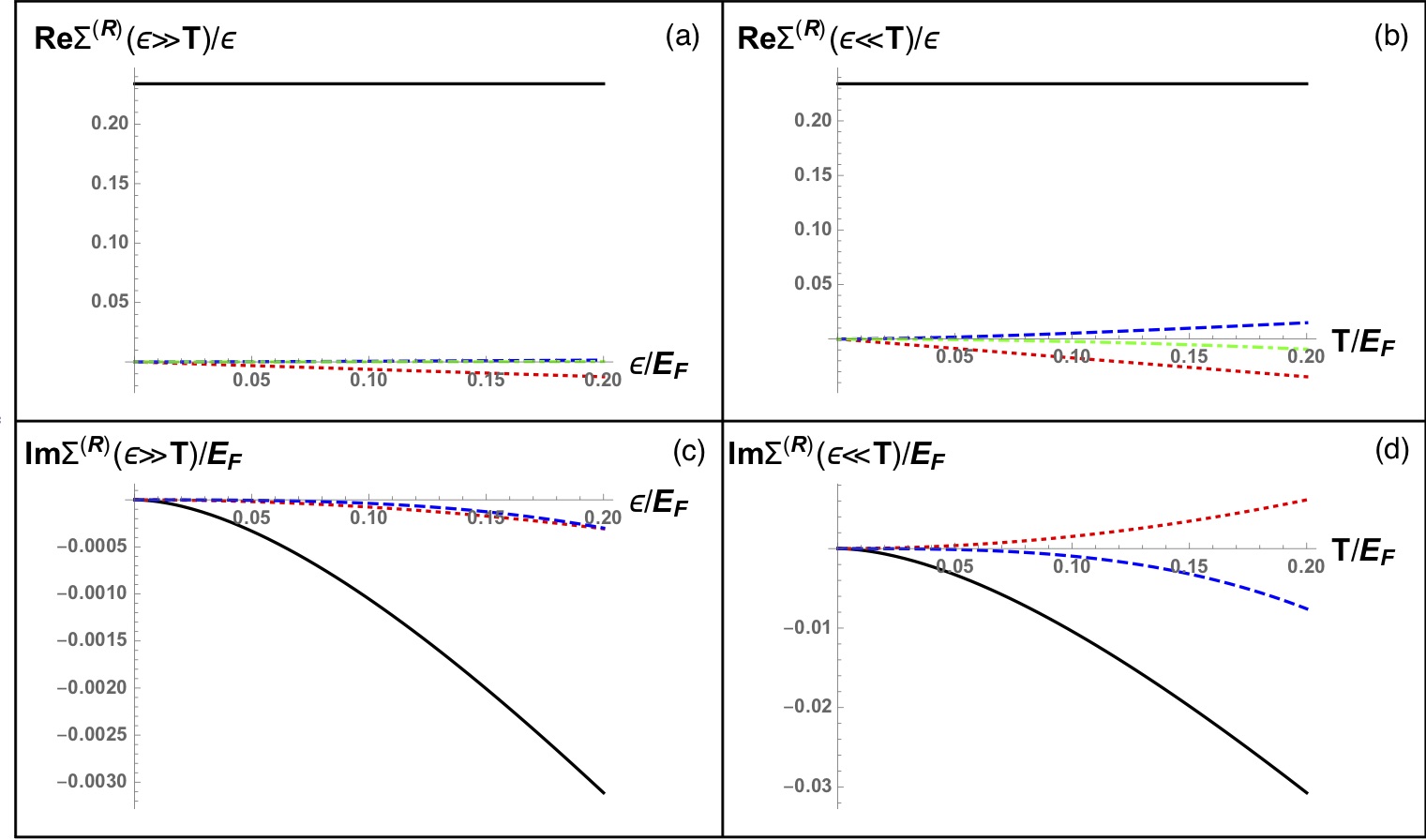}
	\caption{
	Comparison between the leading order term and the higher order terms for
	(a) $\re \Sigma^{(R)}(\e,T)/\e$ in the limit of $\e \gg T$ (Eq.~\ref{eq:ReSig->}), (b) $\re \Sigma^{(R)}(\e,T)/\e$ in the limit of $T \gg \e$ (Eq.~\ref{eq:ReSig-<}), (c) $\im \Sigma^{(R)}(\e,T)$ for $\e \gg T$ (Eq.~\ref{eq:ImSig->}) and (d) $\im \Sigma^{(R)}(\e,T)$  for $T \gg \e$ (Eq.~\ref{eq:ImSig-<}). The leading order self-energy is represented by the black solid curve, while the red dotted, blue dashed and green dash-dotted curves correspond to the remaining terms from lower to higher orders in $\e$ or $T$. $r_s$ is set to $1$ in this figure.
	}
	\label{fig:p4}
\end{figure}

In Fig.~\ref{fig:p4}, the leading order self-energy is compared with the higher order terms, 
 in the limits of $\e/T \ll 1$ and $\e/T \gg 1$. 
 As is apparent in this figure, for low enough energy $\e$ and temperature $T$, the leading term is much larger compared with the remaining higher order terms, and our analytical expressions remain valid as long as this is true.
The explicit range of validity depends on the value of $r_s$.


\section{Conclusion}\label{sec:conclusion}
 
In this paper, we present the calculation of the on-shell ($\e=\xi_{\kb}$) self-energy $\Sigma^{(R)}(\kb,\e)$ of a 2D electron system with Coulomb interactions in the high density limit $r_s \ll 1$.
We work in the regime where the temperature $T$ and the energy $\e$ are arbitrary with respect to each other but both are small and satisfy $r_s^{3/2}  \ll \Delta/\Ef \ll r_s $, $\Delta=\left\lbrace T, |\e| \right\rbrace $.
Thus, we are in the low-energy quasiparticle limit, but we keep the next-to-leading-order terms involving both energy and temperature together.
We obtain analytically the real and imaginary parts of the self-energy for arbitrary values of $\e/T$ up to the order $(\min (|\e|,T)/r_s \Tf)^3$ and to the leading order in $r_s$. 
The asymptotic behaviors of the electron self-energy in the low energy $|\e| \ll T$ and low temperature $T \ll |\e|$ limits are also discussed.

To calculate the electron self-energy, we employ the RPA approximation which is exact in the high-density ($r_s \ll 1$) limit, but is often found to be reliable even outside the high density regime. 
Our results may therefore remain valid even at moderate densities outside the strict high-density limit.
It is well-known that using the noninteracting Green's function $G_0$ to calculate the self-energy instead of the full Green's function $G$ does not change the results in the leading order $r_s$ expansion. The $GW$ approximation and RPA are equivalent to the leading order in $r_s$, but not in higher orders. We believe that RPA is a more consistent approximation because it keeps only the most divergent diagrams in each order whereas $GW$ mixes orders by using the interacting $G$, but at the same time, neglecting vertex corrections. 
Our work is theoretically motivated, providing the analytical expression for the 2D self-energy going beyond the simple leading-order linear in $\varepsilon$ for the real part and quadratic in $T$ (or in $\varepsilon$) for the imaginary part of the self-energy.  The next to the leading order terms involve both energy and temperature combined into nontrivial multiplicative forms, which cannot be guessed from dimensional arguments, showing the subtle and intricate nature of the many body problem even for a system which has been extensively studied for seventy years.  The experimental implications of our work arise in the context of 2D tunneling measurements as carried out in Ref.~\cite{Eisenstein} where the quasiparticle spectral function is measured directly for interacting 2D electrons as a function of temperature and energy (tuned by the bias voltage).  Since the quasiparticle spectral function is determined directly by the real and imaginary parts of the 2D self-energy through the formula that the spectral function is proportional to the imaginary part of the interacting Green's function, such a tunneling spectroscopic measurement can be directly compared with our theory.  We mention that the original experiment was compared with the leading-order theory of Ref.~\cite{Zheng}, and therefore, it will be interesting to compare future such measurements with our analytical theory which goes beyond the leading order self-energy in energy and temperature.  One caveat here is that the typical experimental $r_s$ parameter here is $r_s \sim 1$, which does not satisfy the high-density $(r_s \ll 1)$ RPA requirement.  We note, however, that 3D metals have $r_s \sim 5$, and RPA theories have had great success in describing metallic properties and band structures through the ``$GW$'' approximation, most likely because of the cancellation of higher-order vertex diagrams as discussed in Ref.~\cite{Rice}.  Thus, there is hope that a comparison between our improved analytical 2D self-energy results with future 2D tunneling measurements could lead to a deeper understanding of the Fermi liquid renormalization effects in 2D Coulomb interacting systems.

Before concluding, it may be useful for us to emphasize some of the salient features of our analytical self-energy results.  The leading-order 2D imaginary self-energy is already known to have the $T^2 \ln T$ (for $T \gg |\varepsilon|$) or $\varepsilon^2 \ln \varepsilon$ (for $ \varepsilon \gg T$)  behavior, with the logarithmic part a special 2D feature not arising in 3D systems.   Our work establishes the next-to-leading-order terms going as $O(T^2$ or $\varepsilon^2)$ and $O(T^3$ or $\varepsilon^3)$, respectively in the 2D imaginary self-energy for $T\gg \varepsilon$ or $T \ll \varepsilon$ as the case may be.  No additional logarithms arise in these higher-order terms.  On the Fermi surface, where $\varepsilon =0$, the quasiparticle broadening therefore goes as $O(T^2 \ln T) + O(T^2) + O(T^3)$.   The real part of the 2D self-energy is even more subtle in our theory because of the non-analytical contributions arising from the special form of the 2D polarization bubble (with a kink at $k=\kf$).  In particular, the leading order results is the usual $O(\varepsilon)$ for $T \gg \varepsilon$ and $\varepsilon \gg T$, which provides the quasiparticle effective mass renormalization to the electronic specific heat.  The next-to-the-leading order terms in the real 2D self-energy are $O(\varepsilon T )+ O(\varepsilon T^2 \ln T ) + O(\varepsilon T^2)$ for $T \gg |\varepsilon|$, indicating a linear-in-$T$ correction to the usual specific heat coefficient in violation of the Sommerfeld expansion.  For $\varepsilon \gg T$, the higher order self-energy corrections to the real part of the 2D self-energy go as $O(\varepsilon^2) + O (\varepsilon ^3 \ln \varepsilon) + O(\varepsilon^3)$-- the appearance of the log here is again special to 2D systems.  We also note that the full RPA low-energy and low-temperature self-energy expression derived by us in this work does not suffer from the leading-order logarithmic corrections found in the Hartree-Fock theories~\cite{Setiawan,Li2013}.  The full analytical expressions (when $\varepsilon \sim T$) for the imaginary and real parts of the 2D self-energy, given in Eqs.~\ref{eq:ImSig} and~\ref{eq:ReSig} respectively, are very complex and do not allow for a simple discussion.


\acknowledgements

This work is supported by the Laboratory for Physical Sciences (D.B., M.S., and S.D.S) and the NSF DMR-1613029 (Y.L.).
 	


\appendix

\appendixheaderon

\section{Derivation of the general formulas for the Fermi liquid self-energy}~\label{sec:AppD}

\subsection{Keldysh formalism for interacting electrons}

In this appendix, we derive the self-energy formulas presented in Sec.~\ref{sec:formula} using the Keldysh technique.
We start from the partition function of a 2D system of interacting electrons:
\begin{align}\label{eq:Z}
\begin{aligned}
Z
=&
\int 
\dd \left( \bar{\psi}, \psi \right) 
\,
\exp
\left( 
iS_0+iS_{\msf{int}}
\right),
\\
S_0
=\,&
\intl{\vex{r},\vex{r'},t,t'}
\bar{\psi}(\vex{r},t)
\;
\hat{G}_0^{-1}(\vex{r},t;\vex{r},t')	
\;
\psi(\vex{r'},t'),
\\
S_{\msf{int}}
=\,&
-
{\frac{1}{2}}
\,
\sum_{a = 1,2}
\zeta_a
\intl{t,\vex{r}}
\bar{\psi}_{a,\sigma}(\xb,t) \bar{\psi}_{a,\sigma'}(\xb',t)
\\
&\times
V(\xb-\xb')
\psi_{a,\sigma'}(\xb',t)\psi_{a,\sigma}(\xb,t).
\end{aligned}
\end{align}
Here $\psi_{a,\sigma}(\rb,t)$ is a Grassmann field that carries a spin index $\sigma \in \{\uparrow,\downarrow\}$ and a Keldysh label $a \in \{1,2\}$.
$a=1$ ($2$) indicates the forward (backward) part of the Keldysh contour, and the corresponding sign factor $\zeta_a$ assumes the value of $+1$ ($-1$).
$V(\rb)=e^2/r$ is the bare Coulomb interaction potential,
and $G_0$ denotes the non-interacting Green's function defined on the Keldysh contour: 
\begin{align}
\begin{aligned}
\hat{G}_0(\xb,t;\xb',t')
\equiv
-i \left\langle \T_{\msf{c}} \, \psi(\xb,t) \, \bar{\psi}(\xb',t')\right\rangle_0.
\end{aligned}	
\end{align}
$\T_{\msf{c}}$ is the Keldysh contour ordering operator, and the angular bracket with subscript ``0'' stands for the functional averaging over the non-interacting action.

We then introduce an auxiliary bosonic field $\phi=\left[ \phicl \quad \phiq \right]^{\T} $ to Hubbard-Stratonovich (H.S.) decouple the interaction:
\begin{align}\label{eq:Sint}
\begin{aligned}
&e^{ i S_{\msf{int}} }
=\,
\int 
\dd \phi
\,
\exp
\left[
i 
\intl{\qb,\ww}
\phicl(\qb,\ww)
V^{-1}(q)
\phiq(-\qb,-\ww) 
\right. 
\\
&
-
\frac{i}{\sqrt{2}}
\intl{\kb,\e, \qb,\ww}	
\phicl(\qb,\ww) 
\,
\bar{\psi}(\kb+\qb,\e+\ww) \htau^3 \psi(\kb,\e) 
\\
&
\left. 
-
\frac{i}{\sqrt{2}}
\intl{\kb,\e, \qb,\ww}	
\phiq(\qb,\ww) 
\,
\bar{\psi}(\kb+\qb,\e+\ww) \, \psi(\kb,\e) 
\right].
\end{aligned}
\end{align}
$\hat{\tau}$ here represents the Pauli matrix acting on the Keldysh space.

To further simplify the calculation, we apply the Keldysh rotation to the fermionic field
\begin{align}\label{CoV1}
\begin{aligned}
\psi(\kb,\e) 
\rightarrow \, 
\htau^3 \uk \psi(\kb,\e),
\qquad
\bar{\psi}(\kb,\e) 
\rightarrow \,
\bar{\psi}(\kb,\e) \, \uk^\dagger,
\end{aligned}
\end{align}
where $\uk \equiv (\hat{1} + i \htau^2)/\sqrt{2}$.
After the rotation, the non-interacting fermionic Green's function $\hat{G}_0$ assumes the following structure in the Keldysh space
\begin{align}
\begin{aligned}
\hat{G}_0(\kb,\e)	
=\,
\begin{bmatrix}
G_0^{(R)}(\kb,\e)	 	& G_0^{(K)}(\kb,\e)	
\\
0 	& G_0^{(A)}  (\kb,\e)
\end{bmatrix},
\end{aligned}
\end{align} 
with the components given by Eqs.~\ref{eq:GR} and~\ref{eq:FDT-G}.

The Keldysh rotation is then followed by a distribution function dependent transformation
\begin{align}\label{CoV2}
\begin{aligned}
\psi(\kb,\e) 
\rightarrow \, 
\mf(\e) \, \psi(\kb,\e),
\qquad
\bar{\psi}(\kb,\e) 
\rightarrow \,
\bar{\psi}(\kb,\e) \, \mf(\e),
\end{aligned}
\end{align}
where $\mf(\e)$ acquires the following form in the Keldysh space
\begin{align}\label{eq:MF}
\begin{aligned}
\mf(\e)
=
\begin{bmatrix}
1  	& \tanh \left( \e/2 T \right) 
\\
0 	& -1  
\end{bmatrix}.
\end{aligned}
\end{align} 
After the combined transformation, the bare fermionic Green's function $\hat{G}_0$ becomes diagonal in the Keldysh space and distribution function independent,
\begin{align}\label{eq:g0}
\begin{aligned}
\hat{G}_0(\kb,\e)	
=\,
\begin{bmatrix}
G_0^{(R)}(\kb,\e)	 &0
\\
0 	& G_0^{(A)}  (\kb,\e)
\end{bmatrix},
\end{aligned}
\end{align}
and the partition function is now given by
\begin{align}\label{eq:Zrot}
\begin{aligned}
&Z
=\,
\int 
\dd \left( \bar{\psi},\psi \right) 
\dd \phi
\,
\exp
\left( i S_{\psi}+i S_{\phi}+iS_c \right),
\\
&S_{\phi}
=\,
\intl{\qb,\ww}
\phicl(\qb,\ww) V^{-1}(q)\phiq(-\qb,-\ww),
\\
&S_{\psi}
=\,
\intl{\kb,\e}
\bar{\psi}(\kb,\e)
\left[\e - \xi_{\kb} + i \eta \htau^3  \right]
\psi(\kb,\e),
\\
&S_c
=\,
-\!\!\!\!\!\!\!
\intl{\kb,\kb',\e,\e'}
\!\!\!\!\!\!\!
\frac{\phicl(\kb-\kb', \e - \e')}{\sqrt{2}}
\bar{\psi}(\kb,\e) 
\mf(\e)	 \mf(\e') 
\psi(\kb',\e')
\\
&\,
-\!\!\!\!\!\!\!
\intl{\kb,\kb',\e,\e'}
\!\!\!\!\!\!\!
\frac{\phiq(\kb-\kb', \e - \e')}{\sqrt{2}}
\bar{\psi}(\kb,\e) 
\mf(\e) \htau^1 \mf(\e')
\psi(\kb',\e').
\end{aligned}
\end{align}

\subsection{Dressed propagator for the H.S. field}

It is clear from Eq.~\ref{eq:Zrot} that the bare propagator of the H.S. field is 
\begin{align}\label{eq:D0}
\begin{aligned}
	\hat{D}_0(\qb,\ww)
	\equiv 
	\,&
	-i
	\braket{
	\phi(\qb,\ww) 
	\phi^{\T}(-\qb,-\ww) 
	}_0
	=
	\begin{bmatrix}
	0 & V(q)
		\\
	V(q) & 0
	\end{bmatrix}.
\end{aligned}
\end{align}
To obtain its dressed RPA propagator, we integrate out the fermionic field $\psi$, and arrive at an effective action $iS_{\phi}+\ln \braket{\exp (iS_c)}_{\psi}$. Here the angular bracket with subscript $\psi$ denotes the functional integration over the field $\psi$ with weight $\exp \left( i S_{\psi} \right)$,
\begin{align}
\begin{aligned}
&\braket{\exp (iS_c)}_{\psi}
\equiv \,
\int \dd \left( \bar{\psi},\psi \right)
\exp \left( i S_{\psi} +iS_c \right).
\end{aligned}
\end{align}

To the leading order in the cumulant expansion, which is equivalent to the random phase approximation (RPA), 
$\ln \braket{\exp (iS_c)}_{\psi}\approx \braket{\frac{1}{2}(iS_c)^2}_{\psi}$ and can be expressed as a quadratic form
\begin{align}\label{eq:Sc2}
\begin{aligned}
\braket{\frac{1}{2}(iS_c)^2}_{\psi}
=
-\frac{i}{2} 
\int_{\qb,\ww}
\phi^{\T}(-\qb,-\ww) 
\hat{\Pi}(\qb,\ww)
\phi(\qb,\ww),
\end{aligned}
\end{align}
where the kernel $\hat{\Pi}(\qb,\ww)$ is the self-energy for the H.S. field $\phi$, with components
	\begin{align}
	\begin{aligned}
	&\Pi^{ab}(\qb,\ww)
	=\,
	-i
	\intl{\kb,\e}
	\Tr
	\left\lbrace 
	\left[ 
	\frac{1+\zeta_a}{2}
	+
	\frac{1-\zeta_a}{2}
	\hat{\tau}^1
	\right] 
	\right. 
	\\
	&\times
	\mf (\e+\ww) 
	G_0(\kb+\qb,\e+\ww) 
	\mf (\e+\ww)
	\\
	&\times
	\left. 
	\left[ 
	\frac{1+\zeta_b}{2}
	+
	\frac{1-\zeta_b}{2}
	\hat{\tau}^1
	\right] 
	\mf (\e) 
	G_0(\kb,\e) 
	\mf (\e)
	\right\rbrace.
	\end{aligned}
	\end{align}

Inserting Eqs.~\ref{eq:g0} and~\ref{eq:MF} into the equation above, and using the causality relation
\begin{align}\label{eq:causality}
\begin{aligned}
\intl{\kb,\e}
G_0^{(R)}(\kb+\qb,\e+\ww)G_0^{(R)}(\kb,\e)=0,
\end{aligned}
\end{align}
we find that $\hat{\Pi}(\qb,\ww)$ possesses the standard causality structure of a bosonic self-energy
\begin{align}\label{eq:Pi}
\begin{aligned}
&\hat{\Pi}(\qb,\ww)
=\,
\begin{bmatrix}
0& \Pi^{(A)}(\qb,\ww)
\\
\Pi^{(R)}(\qb,\ww)& \Pi^{(K)}(\qb,\ww)
\end{bmatrix}.
\end{aligned}
\end{align}
Its retarded component is given by Eq.~\ref{eq:Pi2}
which can be further simplified to
\begin{align}\label{eq:Pi4}
\begin{aligned}
\Pi^{(R)} (\qb,\ww)
=\,&
\intl{\kb}
\frac{\tanh(\xi_{\kb+\qb}/2T)-\tanh(\xi_{\kb}/2T)}{\ww+\xi_{\kb}- \xi_{\kb+\qb}+i\eta},
\end{aligned}
\end{align}
and is related to its advanced and Keldysh components through:
\begin{align}\label{eq:Pi3}
\begin{aligned}
&\Pi^{(A)}(\qb,\ww)
=
\left[ \Pi^{(R)}(\qb,\ww)\right]^* ,
\\
&\Pi^{(K)}(\qb,\ww)=\,\left[  \Pi^{(R)}(\qb,\ww) -\Pi^{(A)} (\qb,\ww) \right] 
\coth\left( \frac{ \ww}{2T}\right).
\end{aligned}
\end{align}

We then combine the actions $iS_{\phi}$ (Eq.~\ref{eq:Zrot}) and $\braket{\frac{1}{2}(iS_c)^2}_{\psi}$ (Eq.~\ref{eq:Sc2}), and obtain the dressed propagator for the H.S. field $\phi$ 
\begin{align}\label{eq:D}
\begin{aligned}
	\hat{D}(\qb,\ww)
	\equiv\,&
	-i
		\braket{
		\phi(\qb,\ww) 
		\phi^{\T}(-\qb,-\ww) 
	}
	\\
	=&
	\left[ \hat{D}_0(\qb,\ww)-\hat{\Pi}(\qb,\ww) \right]^{-1}\!\!\!\!\!\!\!.
\end{aligned}
\end{align}
In Fig.~\ref{fig:p1}(b), we show the diagrammatic representation of the Dyson equation above. Red wavy lines with open and solid dots are used to indicate, respectively, the bare propagator $D_0(\qb,\ww)$ and dressed propagator $D(\qb,\ww)$, while the black bubble represents the bosonic self-energy $\Pi(\qb,\ww)$.

Using Eq.~\ref{eq:D0} and~\ref{eq:Pi}, one can easily see that $D(\qb,\ww)$ admits the following form in the Keldysh space 
\begin{align}
\begin{aligned}
	\hat{D}(\qb,\ww)
	=\,&
	\begin{bmatrix}
	D^{(K)}(\qb,\ww)& D^{(R)}(\qb,\ww)
	\\
	D^{(A)}(\qb,\ww)& 0
	\end{bmatrix},
\end{aligned}
\end{align}
in accordance with the causality structure for a bosonic propagator,
and its components are given by Eqs.~\ref{eq:D1-a} and~\ref{eq:D1-b}.

\subsection{Electron self-energy}


The RPA self-energy diagram for the fermionic field $\psi$ is plotted in Fig.~\ref{fig:p1}(a), where the red wavy and black solid lines represent, respectively, the dressed H.S. propagator $D$ and the bare fermionic propagator $G_0$. 
The corresponding self-energy expression is
\begin{widetext}
	\begin{align}\label{eq:Sig0}
	\begin{aligned}
	\hat{\Sigma}(\kb,\e)
	=&
	\frac{i}{2}
	\int_{\qb,\ww}
	D^{(K)}(-\qb,-\ww)
	\mf(\e) 
	\mf(\e+\ww) 
	\hat{G}_0(\kb+\qb,\e+\ww)
	\mf(\e+\ww) 
	\mf(\e) 
	\\
	+\,&
	\frac{i}{2}
	\int_{\qb,\ww}
	D^{(R)}(-\qb,-\ww)
	\mf(\e) 
	\mf(\e+\ww) 
	\hat{G}_0(\kb+\qb,\e+\ww)
	\mf(\e+\ww) 
	\htau^1
	\mf(\e) 
	\\
	+\,&
	\frac{i}{2}
	\int_{\qb,\ww}
	D^{(A)}(-\qb,-\ww)
	\mf(\e) 
	\htau^1
	\mf(\e+\ww) 
	\hat{G}_0(\kb+\qb,\e+\ww)
	\mf(\e+\ww) 
	\mf(\e).
	\end{aligned}
	\end{align}
\end{widetext}

Making use of the causality relation
\begin{align}\label{eq:causality2}
\begin{aligned}
	\intl{\qb,\ww}
	G_0^{(R)}(\kb+\qb,\e+\ww)D_0^{(R)}(\qb,\ww)=0,
\end{aligned}
\end{align}
as well as the FDT relations Eqs.~\ref{eq:FDT-G} and~\ref{eq:FDT-D}, we find that  $\Sigma(\kb,\e)$'s off-diagonal components vanish in the Keldysh space,
\begin{align}
\begin{aligned}
&\hat{\Sigma}(\kb,\e)
=\,
\begin{bmatrix}
\Sigma^{(R)}(\kb,\e) & 0
\\
0& 	\Sigma^{(A)}(\kb,\e)
\end{bmatrix}.
\end{aligned}
\end{align}
The retarded component $\Sigma^{(R)}(\qb,\ww)=\left[ \Sigma^{(A)}(\kb,\e)\right]^* $ is given by Eq.~\ref{eq:Sig1}, which may be rewritten as
\begin{align}\label{eq:Sig2}
\begin{aligned}
&\Sigma^{(R)} (\kb,\e)
=\,
-
2
\intl{\qb,\ww,\ww'}
\im D^{(R)}(\qb,\ww)
\im G_0^{(R)}(\kb+\qb,\e+\ww')
\\
&\times
\frac{1}{\ww'-\ww-i\eta}
\left[ \coth\left( \frac{\ww}{2T}\right) -\tanh\left( \frac{\e+\ww'}{2T}\right) \right].
\end{aligned}
\end{align}
Here we have employed the Kramers-Kr\"{o}nig relation
\begin{align}\label{eq:KK}
f^{(R)}(\kb,\e)
=
\int_{-\infty}^{\infty}
\dfrac{d\e'}{\pi}
\dfrac{\im f^{(R)}(\kb,\e')}{\e'-\e-i\eta},
\end{align}
for both the bosonic propagator $D$ and the fermionic propagator $G_0$.

Inserting the explicit expression for $G_0$ (Eq.~\ref{eq:GR}) into Eq.~\ref{eq:Sig2}, and performing the angular integration, we arrive at
\begin{widetext}
\begin{align}\label{eq:Sig3}
\begin{aligned}
&\Sigma^{(R)} (\kb,\e)
=\,
\frac{m}{\pi k}
\intl{\ww,\ww'}
\int_0^{\infty} d q
\Theta \left( 1-\left| \frac{m\ww'}{kq}\right|\right) 
\dfrac{ \im D^{(R)}(\qb,\ww)}{
	\sqrt{1-\left( \frac{m\ww'}{kq} \right)^2   } 		
}
\dfrac{1}{\ww'-\Delta \e -\ww-i\eta}
\left[ \coth\left( \frac{\ww}{2T}\right) -\tanh\left( \frac{\e+\ww'-\Delta \e}{2T}\right) \right],
\end{aligned}
\end{align}
where $\Delta \e$ is defined in Eq.~\ref{eq:de}.
 Eq.~\ref{eq:ImSig-0} which gives $\im \Sigma^{(R)}(\kb,\e)$ can be deduced directly from Eq.~\ref{eq:Sig3}.


To obtain $\re \Sigma^{(R)}(\kb,\e)$, we first rewrite Eq.~\ref{eq:Sig3} as
	\begin{align}\label{eq:Sig-5}
	\begin{aligned}
	&\Sigma^{(R)} (\kb,\e)
	=\,
	\frac{m}{4 \pi^2 k}
	\int_0^{\infty} d q
	\int_{-\infty}^{\infty} d\ww'
	\tanh\left( \frac{\e+\ww'-\Delta \e}{2T}\right)
	\dfrac{ 1}{
		\sqrt{1-\left( \frac{m\ww'}{kq} \right)^2   } 		
	}
	D^{(A)}(\qb,\ww'-\Delta \e)
	\Theta \left( 1-\left| \frac{m\ww'}{kq}\right|\right) 
	\\
	+&
	\frac{m}{4 \pi^3k}
	\int_{-\infty}^{\infty} d\ww
	\coth\left( \frac{\ww}{2T}\right)
	\int_0^{\infty} d q
	\im D^{(R)}(\qb,\ww)
	\int_{-\frac{kq}{m}}^{\frac{kq}{m}} d\ww'
	\dfrac{1 }{
		\sqrt{1-\left( \frac{m\ww'}{kq} \right)^2   } 		
	}
	\dfrac{1}{\ww'-\Delta \e -\ww-i\eta},
	\end{aligned}
	\end{align}
with the help of the Kramers-Kr\"{o}nig relation (Eq.~\ref{eq:KK}).
It is then straightforward to see that $\re \Sigma^{(R)}(\kb,\e)$ equals the principal integral of Eq.~\ref{eq:Sig-5} and is given by Eq.~\ref{eq:ReSig-0}.


\subsection{Kramers-Kr\"{o}nig relation}

In this subsection, we show that the integral for $\re \Sigma^{(R)}(\kb,\e)$ (Eq.~\ref{eq:ReSig-0}) can also be obtained directly from that of 
$\im \Sigma^{(R)}(\kb,\e)$ (Eq.~\ref{eq:ImSig-0}) via the Kramers-Kr\"{o}nig transformation (Eq.~\ref{eq:KK}). Inserting Eq.~\ref{eq:ImSig-0} into Eq.~\ref{eq:KK}, we have
\begin{align}\label{eq:KK-S}
\begin{aligned}
	&\re \Sigma^{(R)} (\kb,\e)
	=\,
	\frac{m}{4 \pi^3k}
	\int_{-\infty}^{\infty}
	\dfrac{d\e'}{\e'-\e}
	\int_{-\infty}^{\infty} d\ww
	\coth\left( \frac{\ww}{2T}\right)
	\int_{0}^{\infty}  d q
	\dfrac{ \im D^{(R)}(\qb,\ww)}{
		\sqrt{1 -\left[ \frac{m}{kq}\left( \ww + \Delta \e'  \right)\right] ^2   }  }
	\Theta \left( 1-\left| \frac{m(\ww + \Delta \e' )}{kq}\right|\right)
	\\
	&-
	\frac{m}{4 \pi^3 k}
	\int_{-\infty}^{\infty}
	\dfrac{d\e'}{\e'-\e}
	\int_{-\infty}^{\infty} d\ww
	\tanh\left( \frac{\ww+\e'}{2T}\right) 
	\int_{0}^{\infty}  d q
	\dfrac{ \im D^{(R)}(\qb,\ww)}{
		\sqrt{1 -\left[ \frac{m}{kq}\left( \ww + \Delta \e'  \right)\right] ^2   }  }
	\Theta \left( 1-\left| \frac{m(\ww + \Delta \e' )}{kq}\right|\right),
\end{aligned}
\end{align}
where $\Delta \e'  \equiv \e'-\xi_{\kb}-q^2/2m$.
\end{widetext}

One can apply the transformation $\e' \rightarrow \e'-\ww+\xi_{\kb}+q^2/2m$ to the first term in the equation above, which reduces to the real part of the second term in Eq.~\ref{eq:Sig-5}. After integrating out $\e'$, we obtain the second term in the integral for $\re \Sigma^{(R)}(\kb,\e)$ (Eq.~\ref{eq:ReSig-0}).
On the other hand, for the second term in Eq.~\ref{eq:KK-S}, we first shift $\ww$ by $\ww \rightarrow \ww-\e'+\e$ and then make the transformation $\e' \rightarrow -\e'+\e+\ww$, which leads to
\begin{align}\label{eq:ReS-3}
\begin{aligned}
&
\frac{-m}{4 \pi^3 k}
\int_{-\infty}^{\infty} d\ww
\tanh\left( \frac{\ww+\e}{2T}\right) 
\int_{0}^{\infty} \!\!\! d q
\Theta \left( 1-\left| \frac{m(\ww + \Delta \e )}{kq}\right|\right).
\\
&\times 
\dfrac{1 }{
	\sqrt{1 -\left[ \frac{m}{kq}\left( \ww + \Delta \e \right)\right] ^2   }  }
\int_{-\infty}^{\infty} d\e'
\dfrac{\im D^{(R)}(\qb,\e')}{-\e'+\ww}.
\end{aligned}
\end{align}
Using the Kramers-Kr\"{o}nig relation for dressed H.S. propagator $D^{(R)}(\qb,\e)$, it is straightforward to see that the integration over $\e'$ yields a factor of $-\pi \re D^{(R)}(\qb,\ww)$ and the Eq.~\ref{eq:ReS-3} reduces to the first term in Eq.~\ref{eq:ReSig-0}.

\color{black}


\section{Momentum integrals}~\label{sec:momentum}


This appendix is devoted to the evaluation of the integral defined in Eq.~\ref{eq:I}.
For convenience, we introduce the following dimensionless quantities:
	\begin{align*}
	\delta=\frac{\omega}{4\Ef},
	\quad
	x=\frac{q}{2k_{\msf{F}}},
	\quad
	\alpha=\frac{r_s}{\sqrt{2}}.
	\numb\label{eqn:notation}
	\end{align*}
	The integral $I(\ww)$ then reduced to the form
	\begin{align*}
	I&=
	\frac{m}{\pi}
	\int_{x_1}^{x_2}dx\;D^{(R)}(x,\delta)\;\Bigg[1-\bigg(x-\frac{\delta}{x}\bigg)^2\Bigg]^{-1/2}\\
	&=
	\frac{m}{\pi}
	\int_{x_1}^{x_2}dx\;D^{(R)}(x,\delta)\;x\Big[(x_2^2-x^2)(x^2-x_1^2)\Big]^{-1/2},
	\numb\label{eqn:int}
	\end{align*}
	where $x_{1,2} \equiv q_{\pm}/2k_{\msf{F}}$ satisfies the equation $(x-\delta/x)^2=1$, and is given by:
	\begin{align*}
	x_1&=\frac{1}{2}\Big|1-\sqrt{1+4\delta}\,\Big|\approx|\delta|-\delta^2\sgn\delta+O(\delta^3),
	\\
	x_2&=\frac{1}{2}\Big(1+\sqrt{1+4\delta}\Big)\approx 1+\delta+O(\delta^2).
	\numb
	\end{align*}

Using the variables defined in Eq. (\ref{eqn:notation}), $D^{(R)}$ is given by:
	
\begin{equation}
	D^{(R)}(x,\delta)= \nu^{-1}\alpha\,\frac{(x-\alpha\re\Pi_0\nu^{-1})+i(\alpha\im\Pi_0\nu^{-1})}{(x-\alpha\re\Pi_0\nu^{-1})^2+(\alpha\im\Pi_0\nu^{-1})^2}.
	\label{eqn:vr}
\end{equation}
Here $\Pi_0$ is the zero-temperature polarization bubble~\cite{Stern}, and is given by:
\begin{subequations}	\label{eq:Pi0}
	\begin{align*}
	&\re\Pi_0\nu^{-1}=-1\\&+\frac{1}{2x^2}\sgn\bigg(1-\frac{\delta}{x^2}\bigg)\re\sqrt{-(x_2^2-x^2)(x^2-x_1^2)}\\&+\frac{1}{2x^2}\sgn\bigg(1+\frac{\delta}{x^2}\bigg)\re\sqrt{4\delta x^2-(x_2^2-x^2)(x^2-x_1^2)},
	\numb
	\label{eqn:pire}
	\end{align*}
	\\
	\begin{align*}
	&\im\Pi_0\nu^{-1}=
	-
	\frac{1}{2x^2}\re\sqrt{(x_2^2-x^2)(x^2-x_1^2)}
	\\
	&\qquad\qquad\quad
	+
	\frac{1}{2x^2}\re\sqrt{(x_2^2-x^2)(x^2-x_1^2)-4\delta x^2}.
	\numb
	\end{align*}
\end{subequations}	


	For this calculation, we will work in the regime where $|\delta|\ll \alpha\ll 1$. Additionally, we will assume $\alpha\ll \left( |\delta|/\alpha \right)^2 $, and therefore, we will calculate results to leading order in $\alpha$, but to several orders in $|\delta|/\alpha$. Specifically, we note that since $|\delta|$ is smaller than $\alpha$, we need only keep leading order terms in $\delta$. 
	We wish to evaluate integral (\ref{eqn:int}) as an expansion in $\alpha$ and $|\delta|/\alpha$. To do so, it is necessary to expand the integrand in terms of $x/\alpha$, $\alpha/x$, $x_1/x$ or $x/x_2$. These ratios are all small in some regimes of integration, but become large in other regimes, and therefore the integrand cannot be expanded in any of these ratios across the entire region of integration. To proceed, we must divide the region of integration into three subregions, and then expand in terms of the appropriate ratios in each subregion. Thus we define the boundaries between the subregions $l_1$ and $l_2$ such that $x_1\ll l_1\ll \alpha\ll l_2\ll x_2$. Since we will be dividing up the region of integration, for convenience define $I(a,b)$ to be the contribution to the integral $I$ from the interval $(a,b)$.
	This lack of a single small parameter over the whole integration region is the key technical difficulty hindering the analytical evaluation of the self-energy in the next to the leading order, explaining why it has not been achieved in spite of the long history of the subject.
	
	\subsection{Simplification of $\Pi_0$}
	
	Inside the region of integration, the second term of Eq.~\ref{eqn:pire} vanishes, since the radical is purely imaginary for $x$ between $x_1$ and $x_2$. Likewise, the third term also vanishes so long as $x$ is not too close to the endpoints, specifically if $x_1[1+2\delta+O(\delta^2)]<x<x_2[1-2\delta+O(\delta^2)]$. In order to simplify $\re\Pi_0$, we must exclude the portion of the region of integration that lies outside this interval. Additionally, as we shall show later, in order to simplify $\im\Pi_0$, we will need to exclude a slightly larger region around $x_1$. Thus for some $\beta$ between 0 and 1, we define $a_1=x_1[1+O(|\delta|^{1-\beta})]$ and $a_2=x_2[1-O(|\delta|)]$. {We will show that the contribution to the integral from the excluded regions near the endpoints is higher order in $|\delta|$ than the rest of the integral.
	
	Consider the contribution to the integral from the region $x_1<x<a_1$. For this region, we introduce the change of variables $z^2=x^2-x_1^2$. The upper bound of integration becomes $\sqrt{a_1^2-x_1^2}=O(|\delta|^{(3-\beta)/2})$, and the lower bound is zero. We then need to find an upper bound for $|D^{(R)}|$ in this region. 
	The points where $|D^{(R)}|$ becomes the largest are $z\rightarrow 0$ and  $z\sim|\delta|^{3/2}$. When $z\rightarrow 0$, $\im\Pi_0\nu^{-1}\rightarrow 0$, and $\re\Pi_0\nu^{-1}\rightarrow-1+\re\sqrt{\delta}/x_1$. For $z\sim|\delta|^{3/2}$, there is a point where $(x-\alpha\re\Pi_0\nu^{-1})\rightarrow 0$ if $\delta>0$. At this point, $\im\Pi_0\nu^{-1}=O(|\delta|^{1/2})$. For all cases, $|D^{(R)}|\le\nu^{-1}\alpha \,O(\alpha^{-1})$. The integral then becomes:
	
	\begin{align*}
	&I(x_1,a_1)=\int_0^{O(|\delta|^{\frac{3-\beta}{2}})}dz\;D^{(R)}(z)\,\Big[x_2^2-z^2-x_1^2\Big]^{-1/2}\\
	&=O(|\delta|^{\frac{3-\beta}{2}}).
	\numb
	\end{align*}
	
	Now consider the region $a_2<x<x_2$. For this region, we introduce the change of variables $u^2=x_2^2-x^2$. With this transformation, the limits of integration become 0 to $O(|\delta|^{1/2})$. It is easy to see that in this region, $|\Pi_0\nu^{-1}|=O(|\delta|^{1/2})$, and thus $|D^{(R)}|=\nu^{-1}O(\alpha)$. Then the integral becomes:
	
	\begin{align*}
	&I(a_2,x_2)=\int_0^{O(|\delta|^{1/2})}du\;D^{(R)}(u)\,\Big[x_2^2-u^2-x_1^2\Big]^{-1/2}\\
	&=\alpha\;O(|\delta|^{1/2}).
	\numb
	\end{align*}
	
	For $a_1<x<a_2$, both the second and the third term of Eq. (\ref{eqn:pire}) vanish, and so in this region we have exactly:
	
	\begin{equation}
	\re\Pi_0\nu^{-1}=-1.
	\end{equation}
	
	For $a_1<x<l_2$, we again use the variable transformation $z^2=x^2-x_1^2$. The upper bound of this region becomes $\sqrt{l_2^2-\delta^2}=l_2[1+O(\delta^2/l_2^2)]$, and the lower bound is $O(|\delta|^{(3-\beta)/2})$, as discussed previously. Specifically, we note that $\delta x_1^2/z^2$ is of order $O(|\delta|^\beta)$. Then we simplify $\im\Pi_0$ as follows:
	
	\begin{align*}
	&\im\Pi_0\nu^{-1}
	\\&
	=-
	\frac{zx_2}{2(z^2+x_1^2)}\Big[1-\sqrt{1-\frac{4\delta (z^2+x_1^2)}{z^2x_2^2}}+O\Big(\frac{l_2^2}{x_2^2}\Big)\Big]
	\\
	&=-
	\frac{zx_2}{2(z^2+x_1^2)}\times\frac{4\delta (z^2+x_1^2)}{2z^2x_2^2}\Big[1+O\big(|\delta|^\beta\big)\Big]
	\\&=-
	\frac{\delta}{z}\Big[1+O\big(|\delta|^\beta\big)\Big].
	\numb
	\end{align*}
	
	For $l_2<x<a_2$, all that is required for the computation is the following, which can be shown easily:
	
	\begin{equation}
	 \im\Pi_0\nu^{-1}=O\Big(|\delta|^{1/2},\frac{\delta}{l_2}\Big).
	\end{equation}
	
	\subsection{The $\re I$ integral}
	
	For $x$ between $a_1$ and $l_1$, we again use $z^2=x^2-x_1^2$, and $\re D^{(R)}$ is given by:
	\begin{align*}
	&\re D^{(R)}=\nu^{-1}\frac{1+\frac{x}{\alpha}}{(1+\frac{x}{\alpha})^2+\frac{\delta^2}{z^2}+O(|\delta|^{\frac{3\beta-1}{2}})}\\
	&=\nu^{-1}\frac{1+\frac{x}{\alpha}}{1+\frac{\delta^2}{z^2}}\bigg[1-\frac{\frac{2x}{\alpha}+\frac{x^2}{\alpha^2}}{1+\frac{\delta^2}{z^2}}+\bigg(\frac{\frac{2x}{\alpha}}{1+\frac{\delta^2}{z^2}}\bigg)^2+O\Big(\frac{x^3}{\alpha^3}\Big)\bigg].
	\numb
	\end{align*}
	This can be integrated, giving:
	\begin{align*}
	&\re I(a_1,l_1)=
	\nu
	\int_{O(|\delta|^{\frac{3-\beta}{2}})}^{l_1-\delta^2\!/(2l_1)}dz\;\re D^{(R)}(z)\,\Big[1+O(l_1^2)\Big]\\
	&=\alpha
	\bigg[\frac{z}{\alpha}-\frac{z(5\delta^2+z^2)}{2\alpha^2\sqrt{\delta^2+z^2}}-\frac{\delta}{\alpha}\arctan\frac{z}{\delta}\\&\qquad\quad
	+\frac{5\delta^2}{2\alpha^2}\log\big(z+\sqrt{\delta^2+z^2}\big)\bigg]\bigg|_{O(|\delta|^{\frac{3-\beta}{2}})}^{l_1-\delta^2\!/(2l_1)}\\
	&=\alpha
	\bigg[-\frac{\pi|\delta|}{2\alpha}+\frac{l_1}{\alpha}-\frac{7\delta^2}{4\alpha^2}+\frac{5\delta^2}{2\alpha^2}\log\frac{2l_1}{|\delta|}+\frac{\delta^2}{2\alpha l_1}\\&\qquad\quad-\frac{l_1^2}{2\alpha^2}+O\Big(\frac{l_1^3}{\alpha^3},\;\frac{\delta^3}{\l_1^3},\;\frac{|\delta|^{\frac{3-\beta}{2}}}{\alpha},\;|\delta|^{\frac{3\beta-1}{2}}\Big)\bigg].
	\numb
	\label{eqn:reia1l1}
	\end{align*}
	
	For $x$ between $l_1$ and $l_2$, we can expand in $|\delta|/x$ and $x$, but can no longer expand $\re\Pi_0$ in terms of $x/\alpha$, thus we simply use:
	\begin{equation}
	\re D^{(R)}=\nu^{-1}\frac{1+\frac{x}{\alpha}}{(1+\frac{x}{\alpha})^2+\frac{\delta^2}{x^2}+O(|\delta|^{\frac{3\beta-1}{2}},\,\frac{\delta^4}{x^4})},
	\end{equation}
	Integrating gives:
	\begin{align*}
	&\re I(l_1,l_2)
	=\nu
	\int_{l_1}^{l_2}dx\;\re D^{(R)}(x)\,\bigg[1+\frac{\delta^2}{2x^2}+O\Big(\frac{\delta^3}{x^3},x\Big)\bigg]\\
	&=
	\alpha
	\bigg[-\frac{l_1}{\alpha}-\frac{5\delta^2}{2\alpha^2}+\frac{5\delta^2}{2\alpha^2}\log\frac{\alpha}{l_1}-\frac{\delta^2}{2\alpha l_1}+\frac{l_1^2}{2\alpha^2}+\log\frac{l_2}{\alpha}\\&\qquad\quad+O\Big(\frac{l_1^3}{\alpha^3},\;\frac{\delta^3}{\l_1^3},\;\frac{\alpha}{l_2},\;l_2\Big)\bigg].
	\numb
	\end{align*}

	For $x$ between $l_2$ and $1$, we need only the leading order term in $\alpha/x$, and thus we can simply use:
	\begin{equation}
	D^{(R)}(x)=\nu^{-1}\Big[\frac{\alpha}{x}+O\Big(\frac{\alpha^2}{x^2}\Big)\Big].
	\end{equation}
	Then integrating, we obtain:
	\begin{align*}
	&\re I(l_2,a_2)=\int_{l_2}^{1+O(\delta)}dx\;\Big[\frac{\alpha}{x}+O\Big(\frac{\alpha^2}{x^2}\Big)\Big](1-x^2)^{-1/2}\\
	&=\alpha \Big[\log\frac{2}{l_2}+O\Big(\frac{\alpha}{l_2},\;l_2\Big)\Big].
	\numb
	\end{align*}
	
	Therefore, combining these results, we have:
	\begin{align*}
	&\re I=\alpha
	\bigg[\log\frac{2}{\alpha}-\frac{\pi|\delta|}{2\alpha}+\frac{\delta^2}{\alpha^2}\Big(-\frac{17}{4}+\frac{5}{2}\log\frac{2\alpha}{|\delta|}\Big)\\&\qquad\qquad+O\Big(\frac{\delta^3}{\alpha^3},\;\alpha,\;|\delta|^{\frac{1-\beta}{2}},\;|\delta|^{\frac{3\beta-1}{2}}\Big)\bigg],
	\numb
	\end{align*}
	which is equivalent to Eq.~\ref{eq:I_2}.
	
	
	Once we have $\re I$, we can proceed to calculate the real part of self-energy using
			\begin{align}\label{eq:ReS-1}
			&\begin{aligned}
			\re \Sigma^{(R)} (\e)
			=\,
			\int_{-\infty}^{\infty} 
			\frac{d\ww}{2\pi}
			\tanh\left( \frac{\ww+\e}{2T}\right)
			\re I(\ww),
			\end{aligned}
			\end{align}
	which can be rewritten as Eq.~\ref{eq:ReS-0} after a change of variable $\ww \rightarrow -\ww$ for negative $\ww$.
	However, we note that it might seem at first that Eqs.~\ref{eq:ReS-0} and~\ref{eq:ReS-1} give rise to different results.
	Consider as an example the leading order term in $\re \Sigma^{(R)}$ which is proportional to the following integral
	\begin{align}\label{eq:I1}
	\int_{-\infty}^{\infty} d\ww \tanh\left[ ( \ww+ \e )/2T\right]. 
	\end{align}
	Applying the transformation $\ww \rightarrow \ww-\e$ leads to $0$, while rewriting the integral as 
	\begin{align}\label{eq:I2}
	\int_{0}^{\infty} d\ww
	\left\lbrace 
	\tanh\left[ (\ww+\e)/2T \right]
	-
	\tanh \left[ (\ww-\e) /2T \right]
	\right\rbrace ,
	\end{align}
	yields nonvanishing result.
	This discrepancy is due to the fact that integral Eq.~\ref{eq:I1} is not well defined,
	and the accurate way to carry out the integration is to rewrite it as Eq.~\ref{eq:I2} instead of making the shift for the following reasons.
	First of all, $\tanh\left[ ( \ww+ \e )/2T\right] $ has an effective discontinuity at $\ww=-\e$, which can not be simply transferred to $\ww=0$ since ``$\e$'' has a physical meaning.
	Furthermore, when deriving Eq.~\ref{eq:ReS-1}, we performed a small $\ww$ expansion and kept only the leading order terms. This means that the integration possesses a cutoff which leads to a nonzero result after the shift.
	In other words, the small $\ww$ (or $\delta$) expansion is only justified because of the factor $\tanh\left[ ( \ww+ \e )/2T\right] -\tanh\left[ ( \ww- \e )/2T\right] $ which restricts the integration to the region $|\ww|  \lesssim \max(|\e|,T)$.
	
	\subsection{The $\im I$ integral} 
	
	We now consider the imaginary part of $I$. For $a_1<x<l_1$, we use the same transformation $z^2=x^2-\delta^2$ as above. Then we have (to the same order as in the previous section):
	
	\begin{align*}
	&\im I(a_1,l_1)=
	-
	\int_{O(|\delta|^{\frac{3-\beta}{2}})}^{l_1-\delta^2\!/(2l_1)}dz\;\frac{\delta/z}{(\frac{\sqrt{z^2+\delta^2}}{\alpha}+1)^2+(\delta/z)^2}\\
	&=
	-
	\int_{O(|\delta|^{\frac{3-\beta}{2}})}^{l_1-\delta^2\!/(2l_1)}dz\;\frac{\delta}{z}\bigg(1+\Big(\frac{\delta}{z}\Big)^2\bigg)^{-1}\times\\&\qquad\qquad\bigg[1-\frac{2\sqrt{z^2+\delta^2}}{\alpha}\bigg(1+\Big(\frac{\delta}{z}\Big)^2\bigg)^{-1}\bigg]\\
	&=
	-
	{\alpha}
	\bigg[\frac{\delta}{\alpha}\log\frac{l_1}{|\delta|}+4\frac{\delta^2}{\alpha^2}\sgn\delta-2\frac{l_1\delta}{\alpha^2}\bigg].
	\numb
	\end{align*}
	For $l_1<x<l_2$:	
	\begin{align*}
	&\im I(l_1,l_2)=
	-
	\int_{l_1}^{l_2}dx\;\frac{\delta\alpha^2}{x(x+\alpha)^2}\\
	&=
	-
	\alpha
	\bigg[-\frac{\delta}{\alpha}+\frac{\delta}{\alpha}\log\frac{\alpha}{l_1}+2\frac{l_1\delta}{\alpha^2}\bigg].
	\numb
	\end{align*}
	For $l_2<x<a_2$, there is no contribution to zeroth order in $\alpha$, since the integrand is of order $\alpha$. Then we have:
	\begin{equation}
	\im I=
	-
	\alpha
	\bigg[\frac{\delta}{\alpha}\bigg(-\!1+\log\frac{\alpha}{|\delta|}\bigg)+4\frac{\delta^2}{\alpha^2}\sgn\delta\bigg],
	\end{equation}
which leads to Eq.~\ref{eq:ImS-0I}.

\subsection{Higher order terms in $\alpha$}

We also calculate $\re I$ to first order in $\alpha$, but to only first order in $\delta/\alpha$. This means that we must keep terms of order $|\delta|$. Additionally, we also keep second order terms in $\alpha$, but zeroth order in $\delta/\alpha$. 
In other words, we now consider the region where $ T/\Ef$ and $|\e|/\Ef$ are of the order of $r_s^2$, unlike in the previous sections.
Then $I(x_1,a_1)$ is still of higher order; however, we do need to include $I(a_2,x_2)$. Again using $u^2=x_2^2-x^2$, we find:

\begin{align*}
&\re I(a_2,x_2)
= \nu \int_0^{\sqrt{x_2^2-a_2^2}}du
\;D^{(R)}(u)\,\Big[x_2^2-u^2-x_1^2\Big]^{-1/2}\\
&=\alpha\Big[\sqrt{x_2^2-a_2^2}+O\big(\alpha|\delta|^{1/2}\big)\Big].
\numb
\end{align*}

To first order in $\delta/\alpha$, there are no additional terms in $\re I(a_1,l_1)$, and thus we use the result from Eq. (\ref{eqn:reia1l1}) above:

\begin{align*}
\re I(a_1,l_1)
=&\alpha
\bigg[-\frac{\pi|\delta|}{2\alpha}+\frac{l_1}{\alpha}\\&+O\Big(\frac{l_1^2}{\alpha^2},\;\frac{\delta^2}{l_1^2},\;\frac{|\delta|^{\frac{3-\beta}{2}}}{\alpha},\;|\delta|^{\frac{3\beta-1}{2}}\Big)\bigg].
\numb
\end{align*}
For $\re I(l_1,l_2)$, we find:
\begin{align*}
&\re I(l_1,l_2)
= \alpha
\int_{l_1}^{l_2}\frac{dx}{x_2}\;\frac{1}{x+\alpha}\,\bigg[1+\frac{x^2}{2x_2^2}+O\Big(\frac{\delta^2}{x^2},x^3\Big)\bigg]\\
&=
\alpha
\bigg[-\frac{l_1}{\alpha}+\frac{\alpha}{l_2}+(1-\delta)\log\frac{l_2}{\alpha}-\frac{\alpha^2}{2l_2^2}-\frac{\alpha l_2}{2}+\frac{l_2^2}{4}\\&\qquad\quad+\frac{\alpha^2}{2}\log l_2+O\Big(\frac{l_1^2}{\alpha^2},\;\frac{\delta^2}{\l_1^2},\;\frac{\alpha^3}{l_2^3},\;l_2^3\Big)\bigg].
\numb
\end{align*}
Finally, we calculate $\re I(l_2,a_2)$:
\begin{align*}
&\re I(l_2,a_2)
=\alpha
\int_{l_2}^{a_2}dx\;\frac{1}{x}\Big[1-\frac{\alpha}{x}+O\Big(\frac{\alpha^2}{x^2}\Big)\Big](x_2^2-x^2)^{-1/2}\\
&=\alpha
\Big[-\frac{\alpha}{l_2}+(1-\delta)\log\frac{2(1+\delta)}{l_2}-\sqrt{x_2^2-a_2^2}-\frac{\alpha^2}{4}\\&\qquad\quad+\frac{\alpha^2}{2l_2^2}+\frac{\alpha l_2}{2}-\frac{l_2^2}{4}+\frac{\alpha^2}{2}\log\frac{2}{l_2}+O\Big(\frac{\alpha^3}{l_2^3},\;l_2^3\Big)\Big].
\numb
\end{align*}

Thus, including our result in the previous section, to combined order in $\alpha$ and $\delta/\alpha$ no more than 2, $\re I$ is given by:
\begin{align*}\label{eq:Ihigher}
\re I&
=\alpha
\bigg[\log\frac{2}{\alpha}-\frac{\pi|\delta|}{2\alpha}+\delta\bigg(1-\log\frac{2}{\alpha}\bigg)\\&+\frac{\alpha^2}{4}\Big(-1+2\log 2\Big)+\frac{\delta^2}{\alpha^2}\Big(-\frac{17}{4}+\frac{5}{2}\log\frac{2\alpha}{|\delta|}\Big)\bigg].
\numb
\end{align*}
We emphasize that this is only the higher-order in $\alpha$ (i.e. in $r_s$) contribution arising from the ring diagrams (and in fact, just from the $\re I$ integral).  There are other contributions to the higher-order terms in $\alpha$ which are beyond the scope of the current work where our interest is to get the exact leading order result in $\alpha$ (i.e. $r_s$).  We in fact expect the last term in Eq.~\ref{eq:Ihigher} to be canceled by a contribution coming from one of our neglected effects.  We show the result in Eq.~\ref{eq:Ihigher} for the sake of completeness in providing the structure of $\re I$ only, and do not use this form in the main part of our paper where our interest is the leading-order in $r_s$ exact theory.

To obtain the real part of the electron self-energy $\re \Sigma^{(R)}(\e,T)$ when $T/\Ef$ and $|\e|/\Ef$ are of the same order as $r_s^2$, we need to insert Eq.~\ref{eq:Ihigher} into the first term in Eq.~\ref{eq:ReSig-0} and carry out the frequency integration. We also emphasize that the second term in Eq.~\ref{eq:ReSig-0} might be nonvanishing in this case and contributes to $\re \Sigma^{(R)}(\e,T)$ as well. This is clearly beyond the scope of the current work.

\section{Integrals with hyperbolic functions $\tanh$ and $\coth$}\label{sec:integrals}

In this appendix, we evaluate the integrals which appear in the calculation of the electron self-energy, and involve the hyperbolic functions $\tanh (x)$ and $\coth (x)$.

We first consider an integral of the following form
\begin{align}
\begin{aligned}
	I_1(a)\equiv&
	\int_0^{\infty}
	dx
	f(x) 
	\\
	\times &
	\left[ 
	2\coth (x)
	-\tanh (x+a)
	-\tanh (x-a)
	\right].
\end{aligned}	
\end{align}
Expressing the hyperbolic functions in terms of  exponential series as in Eq.~\ref{eq:expser} separately for the regimes $a > x \geq 0$ and $x \geq a$, 
we arrive at
\begin{align}\label{eq:cothf}
\begin{aligned}
	& I_1(a)
	\\
	=&
	2\sum_{k=1}^{\infty}  
	  \int_0^{\infty} dx f(x)e^{-2kx} 
	\left[ 
	2
	-
	(-1)^ke^{-2ka}
	-
	(-1)^k e^{2ka}
	\right] 
	\\
	+&
	4\sum_{k=1}^{\infty}  
	\int_{0}^{a} dx 
	f(x)
	(-1)^k
	\cosh \left( {2kx-2ka} \right) 
	+
	2
	\int_{0}^{a}
	dx
	f(x)
	\\
	=&
	2\sum_{k=1}^{\infty}  
	\int_0^{\infty} dx f(x)e^{-2kx} 
	\left[ 
	2
	-
	(-1)^ke^{-2ka}
	-
	(-1)^k e^{2ka}
	\right] 
\end{aligned}
\end{align}
where in the second equality we have used the fact that $\sum_{k=1}^{\infty} (-1)^k \cosh(2kx-2ka)=-1/2$.

We then evaluate the integral $ \int_0^{\infty} dx f(x)e^{-2kx} $ for different $f(x)$,  insert the result back into Eq.~\ref{eq:cothf}, and then perform the summation. This leads to
\allowdisplaybreaks
\begin{align*}\label{eq:hy-Im}
&
\int_0^{\infty} d x
\left[ 2 \coth\left( x\right) 
-\tanh\left( x+a\right)-\tanh\left( x-a \right) \right] 
x
\\
&=\,
\frac{\pi^2}{4}+a^2,
\\
&\int_0^{\infty} d x
\left[ 2 \coth\left( x\right) 
-\tanh\left( x+a\right)-\tanh\left( x-a \right) \right] 
x\ln x
\\
&=\,
\left( 1- \gamma_E-\ln 2 \right) a^2
+\frac{\pi^2}{12}
\left( 3-\gamma_E-\ln \frac{2}{\pi^2} -24 \ln A \right)
\\
&\,\,\,\,\,\,
-\frac{1}{2}
\left[ \partial_s \Li_s (-e^{-2a})+ \partial_s \Li_s (-e^{2a})\right] \bigg\lvert_{s=2},
\\
&\int_0^{\infty} d x
\left[ 2 \coth\left( x\right) 
-\tanh\left( x+a\right)-\tanh\left( x-a \right) \right] 
x^2
\\
&=\,
\zeta(3)-\frac{1}{2}\left[ \Li_3(-e^{-2a})+  \Li_3(-e^{2a}) \right].
\numb
\end{align*}
Here $\gamma_E$ and $A$ represent, respectively, the Euler's constant and Glaisher's constant.  $\Li_s(z)$ and $\zeta(z)$ denote the polylogarithm function and the Riemann zeta function.

Similarly, making use of the expansion in Eq.~\ref{eq:expser}, the integral
\begin{align}
\begin{aligned}
	I_2(a)
	=
	\int_0^{\infty}
	dx
	f(x)
	\left[ 
	\tanh (x+a)
	-\tanh (x-a)
	\right] ,
\end{aligned}
\end{align}
can be rewritten in the form
\begin{align}
\begin{aligned}
	&I_2(a)
	=\,
	2\sum_{k=1}^{\infty} 
	 \int_0^{\infty} dx f(x)e^{-2kx} 
	(-1)^k
	\left[ 
	e^{-2ka}
	-
	e^{2ka}
	\right].
\end{aligned}
\end{align}
It is then straightforward to show that
\begin{align*}\label{eq:hy-Re}
	&	\int_0^{\infty}
	dx
	\left[ 
	\tanh (x+a)
	-\tanh (x-a)
	\right] 
	\\
	&=\,
	2a,
	\\
	&	\int_0^{\infty}
	dx
	\left[ 
	\tanh (x+a)
	-\tanh (x-a)
	\right] 
	x
	\\
	&=\,
	\frac{1}{2}\left[ \Li_2(-e^{-2a})-\Li_2(-e^{2a}) \right] ,
	\\
	&
	\int_0^{\infty}
	dx
	\left[ 
	\tanh (x+a)
	-\tanh (x-a)
	\right] 
	x^2
	\\
	&=\,
	\frac{2}{3}a^3+\frac{\pi^2}{6}a,
	\\
	&
	\int_0^{\infty}
	dx
	\left[ 
	\tanh (x+a)
	-\tanh (x-a)
	\right] 
	x^2\ln x
	\\
	&=\,
	\left( 3-2\gamma_E-\ln 4 \right) 
	\left( \frac{1}{3}a^3+\frac{\pi^2}{12} a \right)
	\\
	&\,\,\,+
	\frac{1}{2}
	\left[ \partial_s \Li_s (-e^{-2a}) - \partial_s \Li_s (-e^{2a}) \right] |_{s=3}.
	\numb
\end{align*}

\bigskip
\bigskip
\bigskip
\bigskip

\bibliography{SelfEnergy}

\begin{thebibliography}{41}%
\makeatletter
\providecommand \@ifxundefined [1]{%
 \@ifx{#1\undefined}
}%
\providecommand \@ifnum [1]{%
 \ifnum #1\expandafter \@firstoftwo
 \else \expandafter \@secondoftwo
 \fi
}%
\providecommand \@ifx [1]{%
 \ifx #1\expandafter \@firstoftwo
 \else \expandafter \@secondoftwo
 \fi
}%
\providecommand \natexlab [1]{#1}%
\providecommand \enquote  [1]{``#1''}%
\providecommand \bibnamefont  [1]{#1}%
\providecommand \bibfnamefont [1]{#1}%
\providecommand \citenamefont [1]{#1}%
\providecommand \href@noop [0]{\@secondoftwo}%
\providecommand \href [0]{\begingroup \@sanitize@url \@href}%
\providecommand \@href[1]{\@@startlink{#1}\@@href}%
\providecommand \@@href[1]{\endgroup#1\@@endlink}%
\providecommand \@sanitize@url [0]{\catcode `\\12\catcode `\$12\catcode
  `\&12\catcode `\#12\catcode `\^12\catcode `\_12\catcode `\%12\relax}%
\providecommand \@@startlink[1]{}%
\providecommand \@@endlink[0]{}%
\providecommand \url  [0]{\begingroup\@sanitize@url \@url }%
\providecommand \@url [1]{\endgroup\@href {#1}{\urlprefix }}%
\providecommand \urlprefix  [0]{URL }%
\providecommand \Eprint [0]{\href }%
\providecommand \doibase [0]{http://dx.doi.org/}%
\providecommand \selectlanguage [0]{\@gobble}%
\providecommand \bibinfo  [0]{\@secondoftwo}%
\providecommand \bibfield  [0]{\@secondoftwo}%
\providecommand \translation [1]{[#1]}%
\providecommand \BibitemOpen [0]{}%
\providecommand \bibitemStop [0]{}%
\providecommand \bibitemNoStop [0]{.\EOS\space}%
\providecommand \EOS [0]{\spacefactor3000\relax}%
\providecommand \BibitemShut  [1]{\csname bibitem#1\endcsname}%
\let\auto@bib@innerbib\@empty
\bibitem [{\citenamefont {Abrikosov}\ \emph {et~al.}(1975)\citenamefont
  {Abrikosov}, \citenamefont {Gorkov},\ and\ \citenamefont
  {Dzyaloshinski}}]{AGD}%
  \BibitemOpen
  \bibfield  {author} {\bibinfo {author} {\bibfnamefont {A.~A.}\ \bibnamefont
  {Abrikosov}}, \bibinfo {author} {\bibfnamefont {L.~P.}\ \bibnamefont
  {Gorkov}}, \ and\ \bibinfo {author} {\bibfnamefont {I.~E.}\ \bibnamefont
  {Dzyaloshinski}},\ }\href@noop {} {\emph {\bibinfo {title} {Methods of
  quantum field theory in statistical physics}}}\ (\bibinfo  {publisher}
  {Dover, New York},\ \bibinfo {year} {1975})\BibitemShut {NoStop}%
\bibitem [{\citenamefont {Quinn}\ and\ \citenamefont
  {Ferrell}(1958)}]{Quinn1958}%
  \BibitemOpen
  \bibfield  {author} {\bibinfo {author} {\bibfnamefont {J.~J.}\ \bibnamefont
  {Quinn}}\ and\ \bibinfo {author} {\bibfnamefont {R.~A.}\ \bibnamefont
  {Ferrell}},\ }\href@noop {} {\bibfield  {journal} {\bibinfo  {journal} {Phys.
  Rev.}\ }\textbf {\bibinfo {volume} {112}},\ \bibinfo {pages} {812} (\bibinfo
  {year} {1958})}\BibitemShut {NoStop}%
\bibitem [{\citenamefont {Chaplik}(1971)}]{Chaplik}%
  \BibitemOpen
  \bibfield  {author} {\bibinfo {author} {\bibfnamefont {A.}~\bibnamefont
  {Chaplik}},\ }\href@noop {} {\bibfield  {journal} {\bibinfo  {journal} {Sov.
  Phys. JETP}\ }\textbf {\bibinfo {volume} {33}},\ \bibinfo {pages} {997}
  (\bibinfo {year} {1971})}\BibitemShut {NoStop}%
\bibitem [{\citenamefont {Giuliani}\ and\ \citenamefont
  {Quinn}(1982)}]{Quinn1982}%
  \BibitemOpen
  \bibfield  {author} {\bibinfo {author} {\bibfnamefont {G.~F.}\ \bibnamefont
  {Giuliani}}\ and\ \bibinfo {author} {\bibfnamefont {J.~J.}\ \bibnamefont
  {Quinn}},\ }\href@noop {} {\bibfield  {journal} {\bibinfo  {journal}
  {Physical Review B}\ }\textbf {\bibinfo {volume} {26}},\ \bibinfo {pages}
  {4421} (\bibinfo {year} {1982})}\BibitemShut {NoStop}%
\bibitem [{\citenamefont {Zheng}\ and\ \citenamefont
  {Das~Sarma}(1996)}]{Zheng}%
  \BibitemOpen
  \bibfield  {author} {\bibinfo {author} {\bibfnamefont {L.}~\bibnamefont
  {Zheng}}\ and\ \bibinfo {author} {\bibfnamefont {S.}~\bibnamefont
  {Das~Sarma}},\ }\href@noop {} {\bibfield  {journal} {\bibinfo  {journal}
  {Physical Review B}\ }\textbf {\bibinfo {volume} {53}},\ \bibinfo {pages}
  {9964} (\bibinfo {year} {1996})}\BibitemShut {NoStop}%
\bibitem [{\citenamefont {Menashe}\ and\ \citenamefont
  {Laikhtman}(1996)}]{Menashe}%
  \BibitemOpen
  \bibfield  {author} {\bibinfo {author} {\bibfnamefont {D.}~\bibnamefont
  {Menashe}}\ and\ \bibinfo {author} {\bibfnamefont {B.}~\bibnamefont
  {Laikhtman}},\ }\href@noop {} {\bibfield  {journal} {\bibinfo  {journal}
  {Phys. Rev. B}\ }\textbf {\bibinfo {volume} {54}},\ \bibinfo {pages} {11561}
  (\bibinfo {year} {1996})}\BibitemShut {NoStop}%
\bibitem [{\citenamefont {Qian}\ and\ \citenamefont {Vignale}(2005)}]{Vignale}%
  \BibitemOpen
  \bibfield  {author} {\bibinfo {author} {\bibfnamefont {Z.}~\bibnamefont
  {Qian}}\ and\ \bibinfo {author} {\bibfnamefont {G.}~\bibnamefont {Vignale}},\
  }\href@noop {} {\bibfield  {journal} {\bibinfo  {journal} {Phys. Rev. B}\
  }\textbf {\bibinfo {volume} {71}},\ \bibinfo {pages} {075112} (\bibinfo
  {year} {2005})}\BibitemShut {NoStop}%
\bibitem [{\citenamefont {Fujimoto}(1990)}]{Fujimoto}%
  \BibitemOpen
  \bibfield  {author} {\bibinfo {author} {\bibfnamefont {S.}~\bibnamefont
  {Fujimoto}},\ }\href@noop {} {\bibfield  {journal} {\bibinfo  {journal}
  {Journal of the Physical Society of Japan}\ }\textbf {\bibinfo {volume}
  {59}},\ \bibinfo {pages} {2316} (\bibinfo {year} {1990})}\BibitemShut
  {NoStop}%
\bibitem [{\citenamefont {Hodges}\ \emph {et~al.}(1971)\citenamefont {Hodges},
  \citenamefont {Smith},\ and\ \citenamefont {Wilkins}}]{Hodges}%
  \BibitemOpen
  \bibfield  {author} {\bibinfo {author} {\bibfnamefont {C.}~\bibnamefont
  {Hodges}}, \bibinfo {author} {\bibfnamefont {H.}~\bibnamefont {Smith}}, \
  and\ \bibinfo {author} {\bibfnamefont {J.}~\bibnamefont {Wilkins}},\
  }\href@noop {} {\bibfield  {journal} {\bibinfo  {journal} {Physical Review
  B}\ }\textbf {\bibinfo {volume} {4}},\ \bibinfo {pages} {302} (\bibinfo
  {year} {1971})}\BibitemShut {NoStop}%
\bibitem [{\citenamefont {Li}\ and\ \citenamefont {Das~Sarma}(2013)}]{Li2013}%
  \BibitemOpen
  \bibfield  {author} {\bibinfo {author} {\bibfnamefont {Q.}~\bibnamefont
  {Li}}\ and\ \bibinfo {author} {\bibfnamefont {S.}~\bibnamefont {Das~Sarma}},\
  }\href@noop {} {\bibfield  {journal} {\bibinfo  {journal} {Physical Review
  B}\ }\textbf {\bibinfo {volume} {87}},\ \bibinfo {pages} {085406} (\bibinfo
  {year} {2013})}\BibitemShut {NoStop}%
\bibitem [{\citenamefont {Malozovsky}\ \emph {et~al.}(1993)\citenamefont
  {Malozovsky}, \citenamefont {Bose},\ and\ \citenamefont
  {Longe}}]{Malozovsky}%
  \BibitemOpen
  \bibfield  {author} {\bibinfo {author} {\bibfnamefont {Y.~M.}\ \bibnamefont
  {Malozovsky}}, \bibinfo {author} {\bibfnamefont {S.~M.}\ \bibnamefont
  {Bose}}, \ and\ \bibinfo {author} {\bibfnamefont {P.}~\bibnamefont {Longe}},\
  }\href@noop {} {\bibfield  {journal} {\bibinfo  {journal} {Phys. Rev. B}\
  }\textbf {\bibinfo {volume} {47}},\ \bibinfo {pages} {15242} (\bibinfo {year}
  {1993})}\BibitemShut {NoStop}%
\bibitem [{\citenamefont {Jalabert}\ and\ \citenamefont
  {Das~Sarma}(1989)}]{Jalabert}%
  \BibitemOpen
  \bibfield  {author} {\bibinfo {author} {\bibfnamefont {R.}~\bibnamefont
  {Jalabert}}\ and\ \bibinfo {author} {\bibfnamefont {S.}~\bibnamefont
  {Das~Sarma}},\ }\href@noop {} {\bibfield  {journal} {\bibinfo  {journal}
  {Phys. Rev. B}\ }\textbf {\bibinfo {volume} {40}},\ \bibinfo {pages} {9723}
  (\bibinfo {year} {1989})}\BibitemShut {NoStop}%
\bibitem [{\citenamefont {Hu}\ and\ \citenamefont {Das~Sarma}(1993)}]{Hu}%
  \BibitemOpen
  \bibfield  {author} {\bibinfo {author} {\bibfnamefont {B.-K.}\ \bibnamefont
  {Hu}}\ and\ \bibinfo {author} {\bibfnamefont {S.}~\bibnamefont {Das~Sarma}},\
  }\href@noop {} {\bibfield  {journal} {\bibinfo  {journal} {Phys. Rev. B}\
  }\textbf {\bibinfo {volume} {48}},\ \bibinfo {pages} {5469} (\bibinfo {year}
  {1993})}\BibitemShut {NoStop}%
\bibitem [{\citenamefont {Narozhny}\ \emph {et~al.}(2002)\citenamefont
  {Narozhny}, \citenamefont {Zala},\ and\ \citenamefont {Aleiner}}]{dephasing}%
  \BibitemOpen
  \bibfield  {author} {\bibinfo {author} {\bibfnamefont {B.~N.}\ \bibnamefont
  {Narozhny}}, \bibinfo {author} {\bibfnamefont {G.}~\bibnamefont {Zala}}, \
  and\ \bibinfo {author} {\bibfnamefont {I.~L.}\ \bibnamefont {Aleiner}},\
  }\href@noop {} {\bibfield  {journal} {\bibinfo  {journal} {Phys. Rev. B}\
  }\textbf {\bibinfo {volume} {65}},\ \bibinfo {pages} {180202(R)} (\bibinfo
  {year} {2002})}\BibitemShut {NoStop}%
\bibitem [{\citenamefont {Reizer}\ and\ \citenamefont
  {Wilkins}(1997)}]{heterostructure}%
  \BibitemOpen
  \bibfield  {author} {\bibinfo {author} {\bibfnamefont {M.}~\bibnamefont
  {Reizer}}\ and\ \bibinfo {author} {\bibfnamefont {J.~W.}\ \bibnamefont
  {Wilkins}},\ }\href@noop {} {\bibfield  {journal} {\bibinfo  {journal} {Phys.
  Rev. B}\ }\textbf {\bibinfo {volume} {55}},\ \bibinfo {pages} {R7363}
  (\bibinfo {year} {1997})}\BibitemShut {NoStop}%
\bibitem [{\citenamefont {Jungwirth}\ and\ \citenamefont
  {MacDonald}(1996)}]{Jungwirt}%
  \BibitemOpen
  \bibfield  {author} {\bibinfo {author} {\bibfnamefont {T.}~\bibnamefont
  {Jungwirth}}\ and\ \bibinfo {author} {\bibfnamefont {A.~H.}\ \bibnamefont
  {MacDonald}},\ }\href@noop {} {\bibfield  {journal} {\bibinfo  {journal}
  {Phys. Rev. B}\ }\textbf {\bibinfo {volume} {53}},\ \bibinfo {pages} {7403}
  (\bibinfo {year} {1996})}\BibitemShut {NoStop}%
\bibitem [{\citenamefont {Galitski}\ and\ \citenamefont
  {Das~Sarma}(2004)}]{Galitski2004}%
  \BibitemOpen
  \bibfield  {author} {\bibinfo {author} {\bibfnamefont {V.~M.}\ \bibnamefont
  {Galitski}}\ and\ \bibinfo {author} {\bibfnamefont {S.}~\bibnamefont
  {Das~Sarma}},\ }\href@noop {} {\bibfield  {journal} {\bibinfo  {journal}
  {Phys. Rev. B}\ }\textbf {\bibinfo {volume} {70}},\ \bibinfo {pages} {035111}
  (\bibinfo {year} {2004})}\BibitemShut {NoStop}%
\bibitem [{\citenamefont {Das~Sarma}\ \emph {et~al.}(2004)\citenamefont
  {Das~Sarma}, \citenamefont {Galitski},\ and\ \citenamefont {Zhang}}]{DS}%
  \BibitemOpen
  \bibfield  {author} {\bibinfo {author} {\bibfnamefont {S.}~\bibnamefont
  {Das~Sarma}}, \bibinfo {author} {\bibfnamefont {V.~M.}\ \bibnamefont
  {Galitski}}, \ and\ \bibinfo {author} {\bibfnamefont {Y.}~\bibnamefont
  {Zhang}},\ }\href@noop {} {\bibfield  {journal} {\bibinfo  {journal}
  {Physical Review B}\ }\textbf {\bibinfo {volume} {69}},\ \bibinfo {pages}
  {125334} (\bibinfo {year} {2004})}\BibitemShut {NoStop}%
\bibitem [{\citenamefont {Zhang}\ and\ \citenamefont
  {Das~Sarma}(2004)}]{Zhang}%
  \BibitemOpen
  \bibfield  {author} {\bibinfo {author} {\bibfnamefont {Y.}~\bibnamefont
  {Zhang}}\ and\ \bibinfo {author} {\bibfnamefont {S.}~\bibnamefont
  {Das~Sarma}},\ }\href@noop {} {\bibfield  {journal} {\bibinfo  {journal}
  {Physical Review B}\ }\textbf {\bibinfo {volume} {70}},\ \bibinfo {pages}
  {035104} (\bibinfo {year} {2004})}\BibitemShut {NoStop}%
\bibitem [{\citenamefont {Ting}\ \emph {et~al.}(1975)\citenamefont {Ting},
  \citenamefont {Lee},\ and\ \citenamefont {Quinn}}]{Ting}%
  \BibitemOpen
  \bibfield  {author} {\bibinfo {author} {\bibfnamefont {C.}~\bibnamefont
  {Ting}}, \bibinfo {author} {\bibfnamefont {T.}~\bibnamefont {Lee}}, \ and\
  \bibinfo {author} {\bibfnamefont {J.}~\bibnamefont {Quinn}},\ }\href@noop {}
  {\bibfield  {journal} {\bibinfo  {journal} {Physical Review Letters}\
  }\textbf {\bibinfo {volume} {34}},\ \bibinfo {pages} {870} (\bibinfo {year}
  {1975})}\BibitemShut {NoStop}%
\bibitem [{\citenamefont {Vinter}(1975)}]{Vinter}%
  \BibitemOpen
  \bibfield  {author} {\bibinfo {author} {\bibfnamefont {B.}~\bibnamefont
  {Vinter}},\ }\href@noop {} {\bibfield  {journal} {\bibinfo  {journal}
  {Physical Review Letters}\ }\textbf {\bibinfo {volume} {35}},\ \bibinfo
  {pages} {1044} (\bibinfo {year} {1975})}\BibitemShut {NoStop}%
\bibitem [{\citenamefont {Schulze}\ \emph {et~al.}(2000)\citenamefont
  {Schulze}, \citenamefont {Schuck},\ and\ \citenamefont {Van~Giai}}]{Schulze}%
  \BibitemOpen
  \bibfield  {author} {\bibinfo {author} {\bibfnamefont {H.-J.}\ \bibnamefont
  {Schulze}}, \bibinfo {author} {\bibfnamefont {P.}~\bibnamefont {Schuck}}, \
  and\ \bibinfo {author} {\bibfnamefont {N.}~\bibnamefont {Van~Giai}},\
  }\href@noop {} {\bibfield  {journal} {\bibinfo  {journal} {Phys. Rev. B}\
  }\textbf {\bibinfo {volume} {61}},\ \bibinfo {pages} {8026} (\bibinfo {year}
  {2000})}\BibitemShut {NoStop}%
\bibitem [{\citenamefont {Rice}(1965)}]{Rice}%
  \BibitemOpen
  \bibfield  {author} {\bibinfo {author} {\bibfnamefont {T.}~\bibnamefont
  {Rice}},\ }\href@noop {} {\bibfield  {journal} {\bibinfo  {journal} {Annals
  of Physics}\ }\textbf {\bibinfo {volume} {31}},\ \bibinfo {pages} {100 }
  (\bibinfo {year} {1965})}\BibitemShut {NoStop}%
\bibitem [{\citenamefont {Chubukov}\ and\ \citenamefont
  {Maslov}(2003)}]{Chubukov2003}%
  \BibitemOpen
  \bibfield  {author} {\bibinfo {author} {\bibfnamefont {A.~V.}\ \bibnamefont
  {Chubukov}}\ and\ \bibinfo {author} {\bibfnamefont {D.~L.}\ \bibnamefont
  {Maslov}},\ }\href@noop {} {\bibfield  {journal} {\bibinfo  {journal} {Phys.
  Rev. B}\ }\textbf {\bibinfo {volume} {68}},\ \bibinfo {pages} {155113}
  (\bibinfo {year} {2003})}\BibitemShut {NoStop}%
\bibitem [{\citenamefont {Chubukov}\ and\ \citenamefont
  {Maslov}(2004)}]{Chubukov2004}%
  \BibitemOpen
  \bibfield  {author} {\bibinfo {author} {\bibfnamefont {A.~V.}\ \bibnamefont
  {Chubukov}}\ and\ \bibinfo {author} {\bibfnamefont {D.~L.}\ \bibnamefont
  {Maslov}},\ }\href@noop {} {\bibfield  {journal} {\bibinfo  {journal} {Phys.
  Rev. B}\ }\textbf {\bibinfo {volume} {69}},\ \bibinfo {pages} {121102(R)}
  (\bibinfo {year} {2004})}\BibitemShut {NoStop}%
\bibitem [{\citenamefont {Gell-Mann}(1957)}]{Gell-Mann}%
  \BibitemOpen
  \bibfield  {author} {\bibinfo {author} {\bibfnamefont {M.}~\bibnamefont
  {Gell-Mann}},\ }\href@noop {} {\bibfield  {journal} {\bibinfo  {journal}
  {Phys. Rev.}\ }\textbf {\bibinfo {volume} {106}},\ \bibinfo {pages} {369}
  (\bibinfo {year} {1957})}\BibitemShut {NoStop}%
\bibitem [{\citenamefont {Li}\ \emph {et~al.}(2011)\citenamefont {Li},
  \citenamefont {Hwang},\ and\ \citenamefont {Das~Sarma}}]{Li2011}%
  \BibitemOpen
  \bibfield  {author} {\bibinfo {author} {\bibfnamefont {Q.}~\bibnamefont
  {Li}}, \bibinfo {author} {\bibfnamefont {E.~H.}\ \bibnamefont {Hwang}}, \
  and\ \bibinfo {author} {\bibfnamefont {S.}~\bibnamefont {Das~Sarma}},\
  }\href@noop {} {\bibfield  {journal} {\bibinfo  {journal} {Physical Review
  B}\ }\textbf {\bibinfo {volume} {84}},\ \bibinfo {pages} {235407} (\bibinfo
  {year} {2011})}\BibitemShut {NoStop}%
\bibitem [{\citenamefont {Setiawan}\ and\ \citenamefont
  {Das~Sarma}(2015)}]{Setiawan}%
  \BibitemOpen
  \bibfield  {author} {\bibinfo {author} {\bibfnamefont {F.}~\bibnamefont
  {Setiawan}}\ and\ \bibinfo {author} {\bibfnamefont {S.}~\bibnamefont
  {Das~Sarma}},\ }\href@noop {} {\bibfield  {journal} {\bibinfo  {journal}
  {Physical Review B}\ }\textbf {\bibinfo {volume} {92}},\ \bibinfo {pages}
  {235103} (\bibinfo {year} {2015})}\BibitemShut {NoStop}%
\bibitem [{\citenamefont {Galitski}\ \emph {et~al.}(2005)\citenamefont
  {Galitski}, \citenamefont {Chubukov},\ and\ \citenamefont
  {Das~Sarma}}]{Galitski2005}%
  \BibitemOpen
  \bibfield  {author} {\bibinfo {author} {\bibfnamefont {V.~M.}\ \bibnamefont
  {Galitski}}, \bibinfo {author} {\bibfnamefont {A.~V.}\ \bibnamefont
  {Chubukov}}, \ and\ \bibinfo {author} {\bibfnamefont {S.}~\bibnamefont
  {Das~Sarma}},\ }\href@noop {} {\bibfield  {journal} {\bibinfo  {journal}
  {Physical Review B}\ }\textbf {\bibinfo {volume} {71}},\ \bibinfo {pages}
  {201302(R)} (\bibinfo {year} {2005})}\BibitemShut {NoStop}%
\bibitem [{\citenamefont {Mahan}(1981)}]{Mahan}%
  \BibitemOpen
  \bibfield  {author} {\bibinfo {author} {\bibfnamefont {G.~D.}\ \bibnamefont
  {Mahan}},\ }\href@noop {} {\emph {\bibinfo {title} {Many-particle physics}}}\
  (\bibinfo  {publisher} {Plenum Press, New York},\ \bibinfo {year}
  {1981})\BibitemShut {NoStop}%
\bibitem [{\citenamefont {Chubukov}\ and\ \citenamefont
  {Maslov}(2012)}]{Chubukov2012}%
  \BibitemOpen
  \bibfield  {author} {\bibinfo {author} {\bibfnamefont {A.~V.}\ \bibnamefont
  {Chubukov}}\ and\ \bibinfo {author} {\bibfnamefont {D.~L.}\ \bibnamefont
  {Maslov}},\ }\href@noop {} {\bibfield  {journal} {\bibinfo  {journal}
  {Physical Review B}\ }\textbf {\bibinfo {volume} {86}},\ \bibinfo {pages}
  {155136} (\bibinfo {year} {2012})}\BibitemShut {NoStop}%
\bibitem [{\citenamefont {Galitskii}()}]{Galitskii1958}%
  \BibitemOpen
  \bibfield  {author} {\bibinfo {author} {\bibfnamefont {V.~M.}\ \bibnamefont
  {Galitskii}},\ }\href@noop {} {}\bibinfo {howpublished} {J. Exptl. Theoret.
  Phys. (U.S.S.R.) 34, 151-162 (1958) [Sov. Phys. JETP, 34(7), 104
  (1958)]}\BibitemShut {NoStop}%
\bibitem [{\citenamefont {Engelbrecht}\ and\ \citenamefont
  {Randeria}(1990)}]{Engelbrecht}%
  \BibitemOpen
  \bibfield  {author} {\bibinfo {author} {\bibfnamefont {J.~R.}\ \bibnamefont
  {Engelbrecht}}\ and\ \bibinfo {author} {\bibfnamefont {M.}~\bibnamefont
  {Randeria}},\ }\href@noop {} {\bibfield  {journal} {\bibinfo  {journal}
  {Physical review letters}\ }\textbf {\bibinfo {volume} {65}},\ \bibinfo
  {pages} {1032} (\bibinfo {year} {1990})}\BibitemShut {NoStop}%
\bibitem [{\citenamefont {Abrikosov}(1988)}]{Abrikosov}%
  \BibitemOpen
  \bibfield  {author} {\bibinfo {author} {\bibfnamefont {A.~A.}\ \bibnamefont
  {Abrikosov}},\ }\href@noop {} {\emph {\bibinfo {title} {Fundamentals of the
  Theory of Metals}}}\ (\bibinfo  {publisher} {North-Holland, Amsterdam},\
  \bibinfo {year} {1988})\BibitemShut {NoStop}%
\bibitem [{\citenamefont {Ando}\ \emph {et~al.}(1982)\citenamefont {Ando},
  \citenamefont {Fowler},\ and\ \citenamefont {Stern}}]{2DES-Rev}%
  \BibitemOpen
  \bibfield  {author} {\bibinfo {author} {\bibfnamefont {T.}~\bibnamefont
  {Ando}}, \bibinfo {author} {\bibfnamefont {A.~B.}\ \bibnamefont {Fowler}}, \
  and\ \bibinfo {author} {\bibfnamefont {F.}~\bibnamefont {Stern}},\
  }\href@noop {} {\bibfield  {journal} {\bibinfo  {journal} {Rev. Mod. Phys.}\
  }\textbf {\bibinfo {volume} {54}},\ \bibinfo {pages} {437} (\bibinfo {year}
  {1982})}\BibitemShut {NoStop}%
\bibitem [{\citenamefont {Hedin}(1965)}]{Hedin}%
  \BibitemOpen
  \bibfield  {author} {\bibinfo {author} {\bibfnamefont {L.}~\bibnamefont
  {Hedin}},\ }\href@noop {} {\bibfield  {journal} {\bibinfo  {journal} {Phys.
  Rev.}\ }\textbf {\bibinfo {volume} {139}},\ \bibinfo {pages} {A796} (\bibinfo
  {year} {1965})}\BibitemShut {NoStop}%
\bibitem [{\citenamefont {Gell-Mann}\ and\ \citenamefont
  {Brueckner}(1957)}]{Gell-Mann-Brueckner}%
  \BibitemOpen
  \bibfield  {author} {\bibinfo {author} {\bibfnamefont {M.}~\bibnamefont
  {Gell-Mann}}\ and\ \bibinfo {author} {\bibfnamefont {K.~A.}\ \bibnamefont
  {Brueckner}},\ }\href@noop {} {\bibfield  {journal} {\bibinfo  {journal}
  {Phys. Rev.}\ }\textbf {\bibinfo {volume} {106}},\ \bibinfo {pages} {364}
  (\bibinfo {year} {1957})}\BibitemShut {NoStop}%
\bibitem [{\citenamefont {Das~Sarma}\ \emph {et~al.}(1979)\citenamefont
  {Das~Sarma}, \citenamefont {Kalia}, \citenamefont {Nakayama},\ and\
  \citenamefont {Quinn}}]{DS1979}%
  \BibitemOpen
  \bibfield  {author} {\bibinfo {author} {\bibfnamefont {S.}~\bibnamefont
  {Das~Sarma}}, \bibinfo {author} {\bibfnamefont {R.~K.}\ \bibnamefont
  {Kalia}}, \bibinfo {author} {\bibfnamefont {M.}~\bibnamefont {Nakayama}}, \
  and\ \bibinfo {author} {\bibfnamefont {J.~J.}\ \bibnamefont {Quinn}},\
  }\href@noop {} {\bibfield  {journal} {\bibinfo  {journal} {Phys. Rev. B}\
  }\textbf {\bibinfo {volume} {19}},\ \bibinfo {pages} {6397} (\bibinfo {year}
  {1979})}\BibitemShut {NoStop}%
\bibitem [{\citenamefont {Nakamura}\ \emph {et~al.}(1978)\citenamefont
  {Nakamura}, \citenamefont {Watanabe},\ and\ \citenamefont
  {Ezawa}}]{Nakamura}%
  \BibitemOpen
  \bibfield  {author} {\bibinfo {author} {\bibfnamefont {K.}~\bibnamefont
  {Nakamura}}, \bibinfo {author} {\bibfnamefont {K.}~\bibnamefont {Watanabe}},
  \ and\ \bibinfo {author} {\bibfnamefont {H.}~\bibnamefont {Ezawa}},\
  }\href@noop {} {\bibfield  {journal} {\bibinfo  {journal} {Surface Science}\
  }\textbf {\bibinfo {volume} {73}},\ \bibinfo {pages} {258 } (\bibinfo {year}
  {1978})}\BibitemShut {NoStop}%
\bibitem [{\citenamefont {Stern}(1967)}]{Stern}%
  \BibitemOpen
  \bibfield  {author} {\bibinfo {author} {\bibfnamefont {F.}~\bibnamefont
  {Stern}},\ }\href@noop {} {\bibfield  {journal} {\bibinfo  {journal}
  {Physical Review Letters}\ }\textbf {\bibinfo {volume} {18}},\ \bibinfo
  {pages} {546} (\bibinfo {year} {1967})}\BibitemShut {NoStop}%
\bibitem [{\citenamefont {Murphy}\ \emph {et~al.}(1995)\citenamefont {Murphy},
  \citenamefont {Eisenstein}, \citenamefont {Pfeiffer},\ and\ \citenamefont
  {West}}]{Eisenstein}%
  \BibitemOpen
  \bibfield  {author} {\bibinfo {author} {\bibfnamefont {S.~Q.}\ \bibnamefont
  {Murphy}}, \bibinfo {author} {\bibfnamefont {J.~P.}\ \bibnamefont
  {Eisenstein}}, \bibinfo {author} {\bibfnamefont {L.~N.}\ \bibnamefont
  {Pfeiffer}}, \ and\ \bibinfo {author} {\bibfnamefont {K.~W.}\ \bibnamefont
  {West}},\ }\href@noop {} {\bibfield  {journal} {\bibinfo  {journal} {Phys.
  Rev. B}\ }\textbf {\bibinfo {volume} {52}},\ \bibinfo {pages} {14825}
  (\bibinfo {year} {1995})}\BibitemShut {NoStop}%
\end{thebibliography}%

\end{document}